\documentclass[letterpaper]{IEEEtran}
\usepackage{graphicx}
\usepackage{physics}
\usepackage{amsmath}
\usepackage{amssymb}
\usepackage{amsthm}
\usepackage{enumitem}
\usepackage{tikz}
\usepackage{xcolor}
\usepackage{bbm}
\usepackage{bm}
\usepackage{mathtools}
\usepackage[T1]{fontenc}

\mathtoolsset{showonlyrefs}
\allowdisplaybreaks

\usepackage{letltxmacro}
\LetLtxMacro{\originaleqref}{\eqref}
\renewcommand{\eqref}{Eq.~\originaleqref}

\usepackage{stmaryrd}

\usepackage[style=ieee,backend=biber]{biblatex}
\bibliography{ref}

\usepackage{ifthen}

\newif\ifarxiv
\arxivtrue

\newif\ifisit
\isitfalse

\newif\ifextra
\extrafalse

\makeatletter
\newcounter{IEEE@bibentries}
\renewcommand\IEEEtriggeratref[1]{\renewbibmacro{finentry}{\stepcounter{IEEE@bibentries}\ifthenelse{\equal{\value{IEEE@bibentries}}{#1}}
    {\finentry\@IEEEtriggercmd}
    {\finentry}}}
\makeatother

\usetikzlibrary{decorations.pathreplacing,positioning,arrows.meta}
\renewcommand{\epsilon}{\varepsilon}

\newcommand{\msn}{\mkern-1mu}

\newcommand{\RM}{\ensuremath{\textnormal{RM}}}

\newtheorem{theorem}{Theorem}
\newtheorem{prop}[theorem]{Proposition}
\newtheorem{rem}[theorem]{Remark}
\newtheorem{lem}[theorem]{Lemma}
\newtheorem{defn}[theorem]{Definition}

\newcommand{\cC}{\mathcal{C}}
\newcommand{\cD}{\mathcal{D}}

\newcommand{\cG}{\mathcal{G}}

\newcommand{\cS}{\mathcal{S}}

\newcommand{\cX}{\mathcal{X}}

\newcommand{\bX}{\bm{X}}
\newcommand{\bY}{\bm{Y}}

\newcommand{\bc}{\bm{c}}

\newcommand{\bg}{\bm{g}}
\newcommand{\bh}{\bm{h}}

\newcommand{\bs}{\bm{s}}
\newcommand{\bt}{\bm{t}}

\newcommand{\bx}{\bm{x}}

 \newcommand{\obs}{O}
 \newcommand{\obsf}{F}
 \newcommand{\obsg}{G}
\newcommand{\mI}{\mathbb{I}}

\newcommand{\hn}{\mathbb{C}^d}
\newcommand{\hop}{\mathbb{S}}
\newcommand{\psd}{\mathbb{S}_{+}}
\newcommand{\pd}{\mathbb{S}_{++}}
\newcommand{\dop}{\mathbb{D}}

\newcommand{\measo}{M}

\newcommand{\mse}{f}
\newcommand{\mmse}{\mathcal{M}}

\renewcommand{\triangleq}{\coloneqq}
\newcommand{\meas}{(\rho^0)^{\otimes n}}

\newcommand{\measgn}{\rho^{\otimes n}}

\newcommand{\inner}[2]{\ensuremath{\left\langle #1, #2 \right\rangle}}
\newcommand{\innerw}[2]{\ensuremath{\left\langle #1, #2 \right\rangle}_{\meas}}
\newcommand{\innerwgn}[2]{\ensuremath{\left\langle #1, #2 \right\rangle}_{\measgn}}

\newcommand{\normw}[1]{\ensuremath{\left\| #1 \right\|_{\meas}}}

\newcommand{\normwgn}[1]{\ensuremath{\left\| #1 \right\|_{\measgn}}}

\newcommand{\basis}[1]{\Omega_{#1}}
\newcommand{\sym}[1]{S_{#1}}
\newcommand{\symn}{S_n}
\newcommand{\swap}{\mathrm{SW}}
\newcommand{\F}[1]{\widehat{#1}}

\renewcommand{\Pr}{\ensuremath{\mathbb{P}}}

\DeclareMathOperator{\supp}{supp}
\DeclareMathOperator{\linspan}{span}

\IEEEoverridecommandlockouts
\usepackage{authblk}

\begin{document}

\date{}

\title{Reed--Muller Codes on CQ Channels via a \\ New Correlation Bound for Quantum Observables\thanks{This material is based upon work supported by the National Science Foundation under Grants 2106213 and 2120757. A short version of this paper has been presented at International Symposium on Information Theory (ISIT) 2025, Michigan, USA. } }

\author[1,3]{Avijit Mandal}
\author[1,2,3]{Henry D. Pfister}
\affil[1]{Department of Electrical and Computer Engineering, Duke University}
\affil[2]{Department of Mathematics, Duke University}
\affil[3]{Duke Quantum Center, Duke University}

\maketitle

\begin{abstract}
The question of whether Reed--Muller (RM) codes achieve capacity on binary memoryless symmetric (BMS) channels has drawn attention since it was resolved positively for the binary erasure channel by Kudekar et al.\ in 2016. In 2021, Reeves and Pfister extended this to prove the bit-error probability vanishes on BMS channels when the code rate is less than capacity. In 2023, Abbe and Sandon improved this to show the block-error probability also goes to zero. These results rely on the symmetry and nested structure of RM codes.

    In this work, we focus on binary-input symmetric classical-quantum (BSCQ) channels and the Holevo capacity. For a BSCQ, we consider observables that estimate the channel input in the sense of minimizing the mean-squared error (MSE).
    Using an orthogonal decomposition of minimum MSE (MMSE) observables under a weighted inner product, we derive a recursion for the extrinsic MMSE of a code bit.  Consequently, for Reed--Muller code sequences whose rates remain below the Holevo capacity by a fixed positive amount, any prescribed set of $2^{o(\sqrt{\log N})}$ bits can be decoded sequentially with the probability of any error tending to zero.
\end{abstract}

\begin{IEEEkeywords}
Reed--Muller codes, classical--quantum channels, quantum observables, Holevo capacity, quantum hypothesis testing, minimum mean-squared error.
\end{IEEEkeywords}

\section{Introduction}

In 1954, Reed--Muller codes were introduced by Muller~\cite{muller1954application}, and their majority-logic decoding was developed by Reed~\cite{reed1954class}. Since then, Reed--Muller codes have played an important role in coding theory, communication, information theory, and theoretical computer science~\cite{Alon-it05,Abbe-now23}. A central question has been whether Reed--Muller codes achieve capacity on natural families of channels.

For classical binary memoryless symmetric channels, this question has seen substantial recent progress. In 2016, Kudekar et al.\ showed that sequences of Reed--Muller codes whose rates converge to a fixed value in $(0,1)$ achieve capacity on the binary erasure channel~\cite{Kudekar-stoc16,Kudekar-it17}. This result was subsequently leveraged and generalized in several directions~\cite{Sberlo-soda20,Abbe-it20,hkazla-stoc21}. In 2021, Reeves and Pfister proved that Reed--Muller codes achieve capacity on binary memoryless symmetric channels with respect to bit-error probability~\cite{Reeves-arxiv21,Reeves-it23}. In 2023, Abbe and Sandon strengthened this to block-error probability for binary memoryless symmetric channels~\cite{Abbe-focs23}. These results rely on two structural features of Reed--Muller codes: their large symmetry group and their nested projection structure. The simplified and generalized framework in~\cite{Pfister-arxiv25a} is the starting point for the present work.

Coding for classical-quantum channels was first studied by Holevo~\cite{holevo1998capacity} and by Schumacher and Westmoreland~\cite{schumacher1997sending}, who showed that reliable communication is possible up to the Holevo capacity using collective measurements across long codeword blocks. This is analogous to Shannon's channel-coding theorem in that it identifies the fundamental communication limit, but the decoding problem is substantially different. The output of a classical-quantum channel is a quantum state, the optimal measurement may be collective, and different candidate codewords generally correspond to noncommuting density matrices. Thus, even for a classical binary linear code, the extrinsic observation of a single bit is not a scalar likelihood-ratio channel but a binary quantum hypothesis-testing problem.

Explicit code constructions and structured decoders for classical-quantum channels have been studied in several directions. Polar codes give near-explicit constructions approaching the Holevo information with quantum successive-cancellation decoding~\cite{wilde2012polar, guha2012polar}. 
Belief propagation with quantum messages was introduced as a quantum analogue of classical message passing for pure-state classical-quantum channels~\cite{Renes-njp17}, with later work proving optimality on tree factor graphs and developing efficient approximate implementations~\cite{piveteau2022quantum}. More recent work studies belief propagation with quantum messages and paired-measurement belief propagation with quantum messages for symmetric classical-quantum channels and polar constructions~\cite{brandsen2022belief,mandal2024polarcodescqchannels,mandal2026belief,mandal2026qmp}.

The present paper is complementary to this line of work.
Rather than designing a low-complexity decoder, we analyze the bitwise optimal decoding measurement for Reed--Muller codes on binary-input symmetric classical-quantum channels.
The goal is to understand whether the symmetry and nesting mechanisms that underlie Reed--Muller capacity results on classical binary memoryless symmetric channels have a quantum analogue. The main challenge is to replace the classical Boolean decoding function and its Fourier analysis by an operator-valued object that can be compared across nested Reed--Muller codes.

At a high level, the proof follows the same philosophy as recent Reed--Muller code analyses on binary memoryless symmetric channels \cite{Reeves-it23,Pfister-arxiv25a}, but each classical object must be replaced by a quantum analogue. In the classical setting, the extrinsic information for a bit can be represented by a Boolean or real-valued decoding function, and the recursive Reed--Muller argument bounds correlations between \emph{two looks} at this function coming from two nested copies of a shorter Reed--Muller code. In the classical-quantum setting, the corresponding object is a Hermitian observable acting on the extrinsic quantum systems. We choose this observable to minimize mean-squared error in estimating the input bit. Its orthogonal decomposition with respect to a weighted GNS inner product plays the role of the Fourier expansion of a classical decoding function. The new correlation inequality for quantum observables is then the operator analogue of the classical two-look correlation bound.

In this work, we focus on binary-input symmetric classical-quantum channels and show that Reed--Muller codes achieve vanishing bit-error probability below the Holevo capacity in a small-set sense. More precisely, for suitable Reed--Muller code sequences of rate below the Holevo capacity, any set of $2^{o(\sqrt{\log N})}$ coordinates can be decoded sequentially with vanishing error probability. 
This result establishes that the Reed--Muller symmetry-and-nesting arguments survive in the noncommutative classical-quantum setting when the classical decoding function is replaced by a quantum observable.

The proof has the following key steps.
First, an EXIT-area argument uses the gap between the code rate and the Holevo capacity to bound the extrinsic entropy.
This entropy bound is then converted into a nontrivial gap for the mean-squared error of the bitwise extrinsic channel.
Next, we use the Reed--Muller nesting structure to compare the optimal mean-squared-error observables associated with two projected copies of a shorter Reed--Muller code.
This gives a recursive relation for the extrinsic mean-squared error.
Lastly, we convert the mean-squared-error bound into a Helstrom bit-error bound and apply Gao's quantum union bound to obtain the small-set decoding result.
The key technical ingredient is a correlation bound for the mean-squared-error observables.
Its proof uses GNS orthogonal decompositions and the transitive symmetry inherited from the Reed--Muller code.

The main contributions are as follows.
\begin{enumerate}[leftmargin=*]

\item We formulate binary classical--quantum estimation as a mean-squared-error optimization over Hermitian observables, characterize the optimal observable, and relate the resulting minimum mean-squared error to the Helstrom error probability.

\item We establish an EXIT-area identity for transitive codes on binary-input symmetric classical--quantum channels and use it to relate the rate gap below the Holevo capacity to the extrinsic uncertainty of a single coordinate.

\item We develop a weighted GNS orthogonal decomposition for quantum observables and obtain a new correlation inequality for observables with overlapping supports and transitive symmetry.

\item We apply these tools to the Reed--Muller nesting structure to prove an explicit decay bound for the optimal bit-error probability below the Holevo capacity and show that any set of $2^{o(\sqrt{\log N})}$ coordinates can be decoded sequentially with vanishing error probability.

\end{enumerate}

\begin{theorem} \label{informal-main-result}
Consider a binary-input symmetric classical-quantum channel $W$ with Holevo capacity $C>0$.
Then, for any $0<\eta < C$, there are constants $c>0$ and $m_0 \in \mathbb{N}$ such that, for all $m>m_0$, there exists $r\in \mathbb{N}_0$ such that the code $\RM(r,m)$ has rate at least $C-\eta$ while any single bit has bit-error probability
\begin{align*}
P_b \leq e^{-c\eta\sqrt{m}}.
\end{align*}
\end{theorem}
This statement follows from Theorem~\ref{error-rate-exponent} and both results are proven in Section~\ref{sec:achieving vanishing ber}.

Establishing a vanishing block-error probability at all rates below the Holevo capacity remains an important direction for future work.
Such a result would also resolve the long-standing open problem of whether Reed--Muller codes achieve strong secrecy on the binary symmetric channel (BSC).
By classical--quantum channel duality \cite{renes2017duality,Rengaswamy-isit21}, achieving vanishing block-error probability on a binary pure-state channel is dual to strong secrecy for the dual code on the (dual) binary symmetric channel.
Since the dual of a Reed--Muller code is again a Reed--Muller code, proving that Reed--Muller codes achieve vanishing block-error probability on binary pure-state channels (at all rates below capacity) would provide the secrecy component needed to show that nested Reed--Muller codes achieve the secrecy capacity of the binary symmetric wiretap channel.

Existing results show that nested Reed--Muller codes can simultaneously provide reliability and strong secrecy at rates close to, but bounded away from, the secrecy capacity~\cite{Pathegama_2023}.
In fact, by duality, \cite{Pathegama_2023}~provides the current best bound for the pure-state noise threshold of Reed--Muller codes for vanishing block-error probability. 
Thus, a vanishing block-error probability result for Reed--Muller codes (at rates below capacity) on binary pure-state channels would not only advance classical--quantum coding theory, but would also resolve the problem of achieving the binary symmetric wiretap secrecy capacity with nested Reed--Muller codes under the strong-secrecy criterion.

\begin{figure*}[!t]
\centering
\begin{tikzpicture}[
    x=1cm,y=1cm,
    >=Latex,
    flow/.style={->,draw=black!65,line width=0.65pt},
    box/.style={
        rounded corners=2pt,
        draw=black!55,
        line width=0.55pt,
        align=center,
        font=\footnotesize,
        text width=3.05cm,
        minimum height=1.42cm,
        inner sep=4pt
    },
    background/.style={box,fill=gray!16},
    mmse/.style={box,fill=blue!13},
    exit/.style={box,fill=orange!20},
    gns/.style={box,fill=green!15},
    symmetry/.style={box,fill=red!12},
    final/.style={box,fill=violet!14},
    mainresult/.style={box,fill=cyan!18,draw=cyan!55!black,line width=0.75pt},
    sectiontitle/.style={font=\scriptsize\bfseries,text=black!72}
]
\node[background] (nesting) at (4.70,5.70)
    {{\scriptsize\bfseries Background}\\[-1pt]
     RM symmetry and nesting\\
     (Prop.~\ref{prop:rm_prop})};
\node[mmse] (optimizer) at (0.50,5.70)
    {{\scriptsize\bfseries MMSE}\\[-1pt]
     MMSE observable\\
     and BSCQ symmetry\\
     (Lemmas~\ref{mmse-observable} and~\ref{pm-symmetry})};
\node[mmse] (helstrom) at (13.10,5.70)
    {{\scriptsize\bfseries MMSE}\\[-1pt]
     MMSE--Helstrom\\
     bound\\
     (Lemma~\ref{mmse-p-error})};

\node[exit] (exitarea) at (8.90,5.70)
    {{\scriptsize\bfseries EXIT}\\[-1pt]
     EXIT area\\
     identity\\
     (Lemma~\ref{exit-cq})};
\node[exit] (rategap) at (8.90,3.80)
    {{\scriptsize\bfseries EXIT}\\[-1pt]
     Rate gap and\\
     extrinsic entropy\\
     (Lemma~\ref{rate-relation-cq})};

\node[gns] (decomp) at (0.50,3.80)
    {{\scriptsize\bfseries Orthogonal decomposition}\\[-1pt]
     GNS decomposition\\
     and symmetry\\
     (Lemmas~\ref{lem:gns-singular} and~\ref{observable-symmetry})};
\node[gns] (corr) at (0.50,1.90)
    {{\scriptsize\bfseries Orthogonal decomposition}\\[-1pt]
     Observable\\
     correlation bound\\
     (Lemma~\ref{cq-transitiveQ})};
\node[symmetry] (decoder) at (4.70,3.80)
    {{\scriptsize\bfseries Code symmetry}\\[-1pt]
     Transitive\\
     decoder observable\\
     (Lemma~\ref{lem:mmse-observable-transitive-symmetry})};
\node[symmetry] (recursion) at (4.70,1.90)
    {{\scriptsize\bfseries Code symmetry}\\[-1pt]
     Extrinsic MMSE\\
     recursion\\
     (Lemma~\ref{recursive-cq})};

\node[final] (initial) at (13.10,3.80)
    {{\scriptsize\bfseries Final error analysis}\\[-1pt]
     Extrinsic entropy\\
     to MMSE\\
     (Lemma~\ref{mmse-entropy})};
\node[final] (iterate) at (8.90,1.90)
    {{\scriptsize\bfseries Final error analysis}\\[-1pt]
     Recursive\\
     MMSE decay\\
     (Lemma~\ref{error-rate})};
\node[mainresult] (main) at (13.10,1.90)
    {{\scriptsize\bfseries Main result}\\[-1pt]
     Rate below capacity\\
     $P_b\leq e^{-c\eta\sqrt m}$\\
     (Theorems~\ref{informal-main-result} and~\ref{error-rate-exponent})};

\draw[flow] (optimizer) -- (decomp);
\draw[flow] (decomp) -- (corr);
\draw[flow] (nesting) -- (decoder);
\draw[flow] (corr) -- (recursion);
\draw[flow] (decoder) -- (recursion);
\draw[flow] (exitarea) -- (rategap);
\draw[flow] (rategap) -- (initial);
\draw[flow] (helstrom) -- (initial);
\draw[flow] (initial) -- (iterate);
\draw[flow] (recursion) -- (iterate);
\draw[flow] (iterate) -- (main);
\end{tikzpicture}
\caption{Proof flow for the main theorem. Boxes with the same color collect results proved in the same section; arrows indicate the principal logical dependencies.}
\label{fig:proof-flow}
\end{figure*}

\section{Background}

\vspace{-1mm}

\subsection{Notation}

\vspace{-0.5mm}

The natural numbers are denoted by $\mathbb{N}=\left\{ 1,2,\ldots\right\} $ and  $\mathbb{N}_0=\mathbb{N}\cup \{0\}$. For $m\in \mathbb{N}_0$,  we use the shorthand $[m]\coloneqq\left\{ 0,\ldots,m-1\right\} $.
The Galois field with two elements (i.e., the
integers $\{0, 1\}$ with addition and multiplication modulo 2) is denoted by $\mathbb{F}_{2}$. The set of permutations of $N$ elements (i.e., bijective mappings from $[N]$ to $[N]$) is denoted by $\sym{N}$.

For $N$-dimensional vectors with elements in a set $\cX$, we write $\bm{x}=(x_0,x_1,\ldots,x_{N-1}) \in \mathcal{X}^{N}$ and we use $\mathbf{0}$ to denote the all-zero vector of length $N$.
The reordering of a vector $\bx \in \cX^N$ by a permutation $\pi \in \sym{N}$ is denoted by $\pi \bx$ and defined by $(\pi \bx)_i = x_{\pi^{-1}(i)}$.
For an $m$-element index set $A=\{a_0,\dots a_{m-1}\}\subseteq [N]$ with $a_{0}<\dots<a_{m-1}$, we define the subvector $x_{A}=(x_{a_0},\dots,x_{a_{m-1}})\in \mathcal{X}^{m}$.
Similarly, the complementary set of elements is denoted by $\sim A \coloneqq [N]\setminus A$ and the implied subvector is $X_{\sim A}$.

The quantum system of $N$ qudits is denoted by $\bY=(Y_{0},\dots, Y_{N-1})$. For an index set $A\subseteq [N]$, the associated quantum system of $|A|$ qudits is denoted by $Y_{A}$ and the complementary set of quantum systems is denoted by $Y_{\sim A}$. The identity operator on the quantum system $Y_{A}$ is denoted by $\mI_{Y_{A}}$ or $\mI$ if the system is clear from the context.

All entropies are in bits unless otherwise specified.

\vspace{-1mm}

\subsection{Preliminaries}
\vspace{-0.5mm}

 In quantum mechanics, a system with $d$ perfectly distinguishable states is called a \emph{qudit} and represented mathematically by a vector in the Hilbert space $\hn$.
Such a vector $\ket{\psi}\in \hn$ with unit norm is called a \emph{pure state}.
 Let $\mathbb{C}^{d\times d}$ denote the vector space of $d \times d$ complex matrices. Let $\hop^d \subset \mathbb{C}^{d\times d}$ denote the subset of operators (i.e., matrices) mapping $\hn$ to $\hn$ that are Hermitian. Let $\psd^d\subset \hop^d$ and $\pd^d \subset \psd^d$ denote the subset of Hermitian operators mapping $\hn$ to $\hn$ which are positive semidefinite and positive definite, respectively.
Let $\dop^d \subset \psd^d$ denote the set of operators on $\hn$ which are positive semidefinite with unit trace.

In quantum mechanics, any \emph{density matrix} $\rho\in \dop^d$ can be written as
\[\rho=\sum_{i \in [d]}p_{i}\ketbra{\psi_{i}}{\psi_{i}},\]
where $p_i \geq 0$ for $i\in [d]$ and $\{\ket{\psi_i}\}_{i\in[d]}$ is an orthonormal collection of pure states.
This form implicitly defines an ensemble of pure states $\Psi = \{p_i,\ket{\psi_i}\}_{i\in [d]}$ where $p_{i}$ is the probability of choosing the pure state $\ket{\psi_{i}}$.
Special care must be taken when $\rho$ is not full rank.
Since $\rank(\rho) = |\supp(p)|$, where $\supp(p) \coloneqq \{i \in [d] \,|\, p_i \neq 0\}$, the support of $\rho$ is defined to be $\supp(\rho) \coloneqq \linspan(\{\ket{\psi_{i}} \,|\, i \in \supp(p)\})$.
The orthogonal projection $P \!=\! \sum_{i \in \supp(p)}\ketbra{\psi_{i}}{\psi_{i}}$ onto $\supp(\rho)$ satisfies $P\rho = \rho P = \rho$. 
\ifarxiv
The unitary evolution of a quantum pure state $\ket{\psi} \in \hn$ is described by the mapping $\ket{\psi} \mapsto U \ket{\psi}$, where $U\in \mathbb{C}^{d\times d}$ is a unitary.
For the pure state ensemble $\Psi$, this evolution results in the modified ensemble $\Psi'=\{p_{i},U\ket{\psi_{i}}\}_{i\in[d]}$ whose density matrix is \vspace{-1mm}
\begin{align*}
    \rho' = \sum_{i=0}^{d-1} p_{i}U\ketbra{\psi_{i}}{\psi_{i}}U^{\dagger}=U\rho U^{\dagger},
\end{align*}
where $U^\dagger$ is the Hermitian transpose of $U$.
\fi
For an operator $\obs\in  \mathbb{C}^{d\times d} $ with SVD $\obs=U\Sigma V^\dagger$, we have
\[
|\obs| = \sqrt{\obs^{\dagger}\obs} = V \Sigma V^\dagger
\]
and the trace norm is defined by
\begin{align*}
\|\obs\|_1=\Tr(\sqrt{\obs^{\dagger}\obs}) = \sum_{i=0}^{d-1} \sigma_i,
\end{align*}
where $\sigma_i$ is the $i$-th singular value.
For operators $\obs_{1}$ and $\obs_2$, the Hilbert-Schmidt inner product is defined by \vspace*{-1mm}
\begin{align*}
   \inner{\obs_1}{\obs_2}_{\mathrm{HS}} =\Tr(\obs_{1}^{\dagger}\obs_2).
\end{align*}

For a single quantum system $Y$, we use the notation $\obs_{Y}$ and $\rho_{Y}$ to denote an observable in $ \hop$ and a density matrix in $\dop$, respectively.
Similarly, we use $\obs_{\bY}$ and $\rho_{\bY}$ for quantum system with $N$ qudits.
For the index set $A\subseteq [N]$, the associated observables and density matrices are denoted by $\obs_{Y_A}$, $\rho_{Y_A},$ $\obs_{Y_{\sim A}}$, and $\rho_{Y_{\sim A}}$, respectively.
When the setup is clear, we sometimes drop the set notation in the subscript.

\subsection{Reed--Muller Codes}
An $[N,K]$ binary linear code $\cC \subseteq\{0,1\}^N$ is a $K$-dimensional subspace of $\mathbb{F}_2^N$.
    Such a code can be defined as the row space of a generator matrix $G \in \mathbb{F}_2^{K \times N}$ or alternatively as the null space of a parity-check matrix $H \in \mathbb{F}_2^{(N-K)\times N}$. 

Two binary codes $\cC, \cC' \subseteq \mathbb{F}_2^N$ are called \emph{equal} if they have the same set of codewords.
We define the projection of $\cC$ onto a subset $I \subseteq [N]$ of its coordinates by
\[ \cC |_I \coloneqq \{ \bc_I \in \{0,1\}^{|I|} \, | \, \bc \in \cC \}. \]
A code $\cC' \subseteq \cX^{|I|}$ is said to be \emph{nested} inside $\cC$ at $I$ if $\cC|_I = \cC'$.

For a binary code $\cC$, the \emph{automorphism group} $\cG$ is
\[\cG \triangleq \{ \pi\in \sym{N} \,|\, \forall \bc \in \cC, \pi \bc \in \cC \}. \]
The permutation group $\cG$ is \emph{transitive} if, $\forall i,j\in [N]$, there is a $\pi \in \mathcal{G}$ such that $\pi(i)=j$. It is \emph{doubly-transitive} if, for distinct $i,j,k\in [N]$, there is a $\pi \in \mathcal{G}$ such that $\pi(i)=i$ and $\pi(j)=k$.
We say that a code $\cC$ is (doubly) transitive if its automorphism group is (doubly) transitive.

The Reed--Muller code $\RM(r,m)$ is a binary linear code of length $N=2^{m}$.
Its codewords are defined by evaluating the set of binary multilinear polynomials with $m$ variables and total degree at most $r$ at all points in $\{0,1\}^m$.
RM codes have rich symmetry properties. For our proofs, we will use some of the following symmetry and nesting properties~\cite{Reeves-it23}.
\begin{prop}\label{prop:rm_prop}
For the sake of brevity, we list only a few key properties of $\cC = \RM( r,m)$ codes to be used later:
\begin{enumerate}[label=(\alph*)]
\item Symmetry: the code $\cC$ is doubly transitive;
\item Projection: $\cC|_{[2^{m-k}]}=\RM(r,m-k)$ when $k\leq m-r$;
\item Nesting: choosing $A = [2^{m-2}]\setminus \{0\}$, $B=[2^{m-2}]+2^{m-2}$, $C = [2^{m-2}]+2\cdot 2^{m-2}$, and $D = [2^{m-2}]+3\cdot 2^{m-2}$ gives $|A|+1=|B|=|C|=|D|=2^{m-2}$ and (see Figure~\ref{rm-nesting})
\[
\cC'=\cC|_{\{0\} \cup A \cup B}
=\cC''=\cC|_{\{0\}\cup A \cup C}
=
\RM(r,m-1).
\]
\end{enumerate}
\end{prop}
These properties enable us to analyze the bit error probability for bit $0$ simultaneously in terms of $\cC$, $\cC'$, and $\cC''$.
In particular, $\cC'$ and $\cC''$ are length-$2^{m-1}$ projections of the code $\cC$: their extrinsic coordinate sets overlap on $A$ while swapping the equal-sized disjoint blocks $B$ and $C$ leaves the overall construction invariant.
\begin{figure}
    \begin{center}
\scalebox{0.6}{\begin{tikzpicture}[xscale=0.9,yscale=-0.9]

\fill[red] (0,0) rectangle ++(2,1);
\fill[blue] (2,0) rectangle ++(2,1);
\fill[yellow!40!orange] (4,0) rectangle ++(2,1);
\fill[green!40!black] (6,0) rectangle ++(2,1);

\draw [
    thick,
    decoration={
        brace,
amplitude=7pt,
        raise=0.1cm
    },
    decorate
] (0,0) -- (4,0)
node  [pos=0.5,anchor=north,yshift=1cm] {$\cC'$}; 

\draw [
    thick,
    decoration={
        brace,
amplitude=7pt,
        raise=0.1cm
    },
    decorate
] (1,-1.35) -- (5,-1.35)
node  [pos=0.5,anchor=north,yshift=1cm] {$\cC''$}; 
\draw [
    thick,
    decoration={
        brace,
amplitude=7pt,
        raise=0.1cm
    },
    decorate
] (0,-1.1) -- (2,-1.1);
\draw [
    thick,
    decoration={
        brace,
amplitude=7pt,
        raise=0.1cm
    },
    decorate
] (4,-1.1) -- (6,-1.1);

\draw [
    thick,
    decoration={
        brace,
amplitude=7pt,
        raise=1.1cm
    },
    decorate
] (0,-1.2) -- (8,-1.2)
node  [pos=0.5,anchor=north,yshift=2cm] {$\cC$};

\draw [
    thick,
    decoration={
        brace,
        mirror,
		amplitude=5pt,
        raise=1cm
    },
    decorate
] (4,0) -- (6,0)
node [pos=0.5,anchor=north,yshift=-1.25cm] {\large $C$}; 

\draw [
    thick,
    decoration={
        brace,
        mirror,
		amplitude=5pt,
        raise=1cm
    },
    decorate
] (6,0) -- (8,0)
node [pos=0.5,anchor=north,yshift=-1.25cm] {\large $D$}; 

\draw [
    thick,
    decoration={
        brace,
        mirror,
		amplitude=5pt,
        raise=1cm
    },
    decorate
] (2,0) -- (4,0)
node [pos=0.5,anchor=north,yshift=-1.25cm] {\large $B$}; 

\draw [
    thick,
    decoration={
        brace,
		mirror,
		amplitude=5pt,
        raise=1cm
    },
    decorate
] (0,0) -- (2,0)
node [pos=0.5,anchor=north,yshift=-1.25cm] {\large $\{0\} \!\cup\! A$}; 
\end{tikzpicture} }
\end{center}
\vspace{-2mm}
    \caption{Diagram showing the nesting structure of $\cC$, $\cC'$, and $\cC''$ with relative sizes shown for the $\RM$ code sequence.}
   \label{rm-nesting}
   \vspace{1mm}
\end{figure}

\subsection{Symmetric Binary-Input Classical-Quantum Channels}

A classical-quantum (CQ) channel $W \colon \cX \to \dop^d$ defined by $x \mapsto \rho^x$ takes a classical binary input $x\in \mathcal{X}$ and produces a quantum state represented by the density matrix $\rho^x$. For the classical input distribution $p_{\mathcal{X}}(x)$, the associated classical quantum state $\rho_{XY}$ is denoted by 
\begin{align*}
    \rho_{XY}=\sum_{x\in \cX} p_{\mathcal{X}}(x) \big(\ketbra{x}{x}_{X}\otimes \rho^{x}_{Y} \big).
\end{align*}
For a classical input $X$ and a quantum output $Y$, the quantum mutual information (or Holevo information) is given by
\vspace*{-0mm}
\begin{align*}
    I(X;Y)_{\rho_{XY}} = S\left(\sum_{x\in \cX}p_{\mathcal{X}}(x)\rho^{x}\right)-\sum_{x\in \cX}p_{\mathcal{X}}(x)S(\rho^x),
\end{align*}
where $S(\rho)=-\Tr(\rho\log_2 \rho)$ is the von Neumann entropy.
\begin{defn}
A binary-input CQ channel $W\colon\{0,1\}\rightarrow \dop^d$ is defined by $x \mapsto \rho^x$.
If there is a unitary $U$ satisfying $U^{2}=\mathbb{I}$ such that $\rho^1=U\rho^0U^\dagger$, then we call this a binary-input symmetric CQ (BSCQ) channel. 
\end{defn}

For the BSCQ channel $W$ with quantum output states $\rho^0$ and $\rho^1$ and input distribution $(1-p,p)$, the Holevo information $I(W)$, is \vspace*{-0.5mm}
denoted by \begin{align*}
    I(W) \coloneqq I(X;Y)_{\rho_{XY}}=S((1-p)\rho^0+p\rho^1)-S(\rho^0)
\end{align*}
where we used the fact that the von Neumann entropy is unitarily invariant, i.e., $S(\rho^1)=S(\rho^0)$.

Since the channel is symmetric, the capacity is achieved at $p=1/2$ and thus the Holevo capacity of the BSCQ channel is given by $C=I(W)|_{p=1/2}$.
Likewise, the channel entropy $H(W)$ is denoted by
\begin{align*}
    H(W) &\ \coloneqq  H(X|Y)_{\rho_{XY}}\\
    &\ =H(X)_{\rho_{XY}}+H(Y|X)_{\rho_{XY}}-H(Y)_{\rho_{XY}}\\
    &\ =h_{2}(p)-I(W),
\end{align*}
where $h_{2}(p)=-p\log_2 p-(1-p)\log_2(1-p)$ is the binary entropy function.
Similarly, the conditional min-entropy is denoted by \vspace*{-2mm}
\begin{align*}
    H_{\min}(W) &\ = H_{\min}(X|Y)_{\rho_{XY}}\\
   &\ =-\log_2\left(\frac{1}{2}+\frac{1}{2} \left\| (1-p)\rho^{0}-p\rho^{1} \right\|_1 \right)
\end{align*}

Throughout, we first discuss the case where $\rho^0$ and $\rho^1$ are full rank and then discuss what adjustments are required for the rank deficient case.

\begin{defn}
An $m$-outcome \emph{projective measurement} of a quantum system in $\mathcal{H}_n$ is defined by a set of $m$ orthogonal projection matrices $\Pi_{j} \in \mathbb{C}^{n\times n}$ satisfying $\Pi_{i}\Pi_{j}=\delta_{i,j}\Pi_{i}$ and $\sum_{j}\Pi_{j}=\mathbb{I}_{n}$, where $\mathbb{I}_n$ is the $n\times n$ identity matrix.
We denote such a measurement by $\hat{\Pi}=\{\Pi_{j}\}_{j=1}^{m}$.
\end{defn}

Applying the measurement $\hat{\Pi}$ to the quantum state $\rho$ results in a random outcome $J$ and the probability of the event $J=j$ is given by $\text{Tr}(\Pi_{j}\rho)$.
The post-measurement state, conditioned on the event $J=j$, is given by $\Pi_{j}\rho\Pi_{j}/ \text{Tr}(\Pi_{j}\rho)$.

Consider a hypothesis test to distinguish between $m$ possible quantum states defined by $ \Phi = \{p_j,\rho_{j}\}_{j=1}^{m}$, where the $j$-th hypothesis has prior probability $p_j$ and density matrix $\rho_j$.
For a projective measurement $\hat{\Pi}$, where $\Pi_j$ is associated with hypothesis $\rho_j$, the probability of choosing correctly is \vspace{-0.5mm}
\begin{align*}
    P(\Phi,\hat{\Pi}) =\sum_{j=1}^m p_{j}\text{Tr}(\Pi_{j}\rho_{j}). \vspace*{-0.5mm}
\end{align*}

\begin{defn} [Helstrom measurement]
Consider the measurement that distinguishes between two quantum states $\rho^{0},\rho^{1}\in\dop^d$ when $\rho^0$ has prior  probability $1-p$.
This measurement minimizes the error probability~\cite{helstrom1969quantum} by using projection operators onto the eigenspaces of \[D=(1-p)\rho^{0}-p\rho^{1}.\]
The resulting ``Helstrom error probability'' is denoted by
\begin{align*}
    P_{e}(\rho^0,\rho^1,p) \coloneqq \frac{1}{2}-\frac{1}{2}\left\|(1-p)\rho^0-p\rho^1 \right\|_{1}.
\end{align*}

Let \(\Pi_+\), \(\Pi_-\), and \(\Pi_\circ \) be the spectral projectors of \(D\) for its positive, negative, and zero eigenspaces.  Every optimal binary projective Helstrom measurement has the form \(\Pi_0=\Pi_+ +Q\) and \(\Pi_1=\Pi_- +(\Pi_\circ -Q)\), where \(Q\) is any orthogonal projector with \(Q\leq\Pi_\circ \).
All such choices have the same average error.
When symmetry between the measurements for the $0/1$ hypotheses is needed, one can use instead the optimal two-outcome POVM operators \(E_0=\Pi_+ +\Pi_\circ/2\) and \(E_1=\Pi_- +\Pi_\circ/2\).
\end{defn}
From \cite{tomamichel2015quantum}, we know that the Helstrom error probability is also related to the min-entropy via
 \begin{align*}
     P_{e}(\rho^0,\rho^1,p)=1-2^{-H_{\min}(W)}.
 \end{align*}
For the uniform prior, this relation gives
\begin{align*}
    P_{e} \left(\rho^0,\rho^1,\frac{1}{2} \right)=\frac{1}{2}-\frac{1}{2}T(\rho^{0},\rho^{1}),
\end{align*}
where $T(\rho^0,\rho^1) \coloneqq \frac{1}{2} \left\|\rho^0-\rho^1 \right\|_1$ is the trace distance.

For a linear code $\cC\subseteq \mathbb{F}_{2}^N$, consider a linear functional $l\colon \cC\rightarrow \mathbb{F}_2$ that is not always 0 and define the following sets $\cC^l_0=\{\bc\in \cC\colon l(\bc)=0\}$ and $\cC^l_1=\{\bc\in \cC\colon l(\bc)=1\}$. Observe that if $\bc,\bc'\in \cC^{l}_0$, then 
\begin{align*}
    l(\bc+\bc')=l(\bc)+l(\bc')=0.
\end{align*}
Thus $\bc+\bc'\in \cC^{l}_0$, which implies $\cC^{l}_0$ is a linear subspace of code $\cC$, hence a subcode.  Since the functional $l$ is nonzero, $\cC^l_1$ is an affine coset of $\cC^l_0$.
 The following properties hold for $\cC^l_0$ and $\cC^l_1$:
\begin{enumerate}
    \item $\cC=\cC^l_0\bigsqcup \cC^l_1$,
    \item $\cC^{l}_0\cap \cC^l_1=\emptyset$,
    \item $|\cC^l_0|=|\cC^l_1|=\frac{|\cC|}{2}$.
\end{enumerate}
A standard example is the linear function $l(\bc) = c_0$ that returns the first bit in the codeword.

Now, we establish a symmetry result that is the CQ analogue of the classical symmetry satisfied by linear codes on symmetric channels.
For $\bc\in \cC$, we define the unitary 
\begin{align*}
    U^{\bc}=\bigotimes_{i=0}^{N-1}U^{c_i}.
\end{align*}
Since $U^{2}=\mI$, we have $(U^{\bc})^{\dagger}=U^{\bc}$ for all $\bc\in \cC$.
The codewords of $\cC$ are transmitted over a BSCQ channel to give $W\colon x\rightarrow U^{x}\rho^0 U^x$ with $x\in \{0,1\}$. 

We assume a uniform distribution on the codewords of $\cC$ and will show that conditioning on the value of the linear functional $l$ gives an average state that is invariant under the $U^{\bc}$ action for all $\bc \in \cC^{l}_{0}$.
Consequently, the corresponding Helstrom projector can be chosen invariant under this action, and expectations over the coset can be evaluated with respect to the all-zero codeword.

Consider the task of discriminating the transmitted codewords on BSCQ channel $W$ based on whether $\bc\in \cC^l_0$ or $\bc\in \cC^l_1$. Then we construct the average density matrices $\sigma^l_x$ for $x\in \{0,1\}$ as below
\begin{align*}
    \sigma^l_x= \frac{2}{|\cC|}\sum_{\bc\in \cC^{l}_x}U^{\bc}(\rho^0)^{\otimes N}U^{\bc}.
\end{align*}

\begin{lem}\label{lem:helstrom-code-symmetry}
Let \(\Pi_+^l,\Pi_-^l,\Pi_\circ^l\) be the positive, negative, and zero spectral projectors for \(\sigma_0^l-\sigma_1^l\).
For every \(\bh\in\cC_0^l\) and \(s\in\{+,-,\circ\}\), \(U^{\bh}\Pi_s^lU^{\bh}=\Pi_s^l\).  For every \(\bg\in\cC_1^l\), \(U^{\bg}\Pi_+^lU^{\bg}=\Pi_-^l\), \(U^{\bg}\Pi_-^lU^{\bg}=\Pi_+^l\), and \(U^{\bg}\Pi_\circ^lU^{\bg}=\Pi_\circ^l\).  Moreover, if an observable \(F\) is invariant under conjugation by every \(U^{\bh}\), \(\bh\in\cC_0^l\), then \(\Tr(F\sigma_0^l)=\Tr(F(\rho^0)^{\otimes N})\).  The same statements hold after puncturing coordinates, with the unitaries and product state restricted to the retained output systems.
\end{lem}

\begin{proof}
    Observe that for $\bc'\in \cC^l_0$
    \begin{align*}
        U^{\bc'}\sigma^l_0 U^{\bc'} & =\frac{2}{|\cC|}\sum_{\bc\in \cC^l_0}U^{\bc'}U^{\bc}(\rho^0)^{\otimes N}U^{\bc}U^{\bc'}\\
        & = \frac{2}{|\cC|}\sum_{\bc\in \cC^l_0}U^{\bc'+\bc}(\rho^0)^{\otimes N}U^{\bc'+\bc}\\
        & = \frac{2}{|\cC|}\sum_{\bc''\in \cC^l_0}U^{\bc''}(\rho^0)^{\otimes N}U^{\bc''}\\
        & = \sigma^l_0.
    \end{align*}
    Similarly $U^{\bc'}\sigma^l_1U^{\bc'}=\sigma^l_1$ for $\bc'\in \cC^l_0$ since $\cC^l_1$ is an affine coset of $\cC^l_0$. Therefore for all $\bc'\in \cC^l_0$, $U^{\bc'}$ commutes with $\sigma^l_0-\sigma^l_1$, since
    \begin{align*}
        U^{\bc'}(\sigma^l_0-\sigma^l_1)U^{\bc'}=\sigma^l_0-\sigma^l_1.
    \end{align*}
    By definition $\Pi^l_+$ is the orthogonal projector to the positive eigenspace of $\sigma^l_0-\sigma^l_1$. Since $U^{\bc'}$ commutes with $\sigma^l_0-\sigma^l_1$ for all $\bc'\in \cC^l_0$, $U^{\bc'}$ preserves the positive eigenspace. Thus 
    \begin{align*}
        U^{\bc'}\Pi^{l}_+U^{\bc'}=\Pi^l_+,\quad \forall \bc'\in \cC^l_0.
    \end{align*}
    Hence this yields 
    \begin{align*}
        \Tr\left(\Pi^l_+\sigma^l_0\right) & =\frac{2}{|\cC|} \sum_{\bc\in \cC^l_0}\Tr\left(U^{\bc}\Pi^l_+U^{\bc}(\rho^0)^{\otimes N}\right)\\
        & = \Tr\left(\Pi^l_+(\rho^0)^{\otimes N}\right).
    \end{align*}
    If \(\bc'\in\cC_1^l\), translation by \(\bc'\) interchanges the cosets $\cC_0^l$ and $\cC_1^l$, so \(U^{\bc'}\sigma_0^lU^{\bc'}=\sigma_1^l\) and \(U^{\bc'}\sigma_1^lU^{\bc'}=\sigma_0^l\).  Hence \(U^{\bc'}(\sigma_0^l-\sigma_1^l)U^{\bc'}=-(\sigma_0^l-\sigma_1^l)\), which swaps the positive and negative spectral projectors and preserves the zero spectral projector.
    Finally, if \(F\) is invariant under translation by $\bh \in \cC_0^l$, then
    \[ \Tr(F\sigma_0^l)=\frac{2}{|\cC|}\sum_{\bc\in\cC_0^l}\Tr(U^{\bc}FU^{\bc}(\rho^0)^{\otimes N})=\Tr(F(\rho^0)^{\otimes N}).\] 
    Restricting every tensor product and unitary to a punctured output system gives the punctured statement.
\end{proof}

\section{Minimum Mean-Squared Error (MMSE) for Quantum Binary Hypothesis Testing}\label{sec:mmse}

In this section, we introduce an estimation formulation of binary CQ hypothesis testing that will be used later in the proof.
For a binary CQ state, the MMSE is obtained by optimizing over Hermitian observables on the quantum output system.  This is useful for two reasons.  First, the optimal observable can be expanded in the GNS basis introduced in later sections.  Second, the resulting MMSE can be compared with the Helstrom error probability, which is the bit-error probability for the induced binary hypothesis test.

\subsection{MMSE Achieving Observable}

For a binary-input CQ channel \(W\) with input distribution $(1-p,p)$, we combine the input and output into the binary CQ state
\begin{align*}
    \rho_{XY} = (1-p) \ket{0}\bra{0}_{ X} \otimes \rho^0_{Y} + p \ket{1}\bra{1}_{X}\otimes \rho^{1}_{Y} ,
\end{align*}
where $\rho^0,\rho^1 \in \dop^{d}$.
A quantum observable on $Y$ is defined by a Hermitian matrix $\measo_Y$ and represents the process of projectively measuring the $Y$-state in the eigenbasis of $\measo_Y$ and returning the eigenvalue associated with the outcome.
Let
\[
\obs_X = (\ket{0}\bra{0}_X - \ket{1}\bra{1}_X)\otimes \mI_Y
\]
be the quantum observable associated with the binary input, and let
\[
\obs_Y = \mI_X \otimes \measo_Y
\]
be any Hermitian observable supported within $Y$.
Optimizing over Hermitian $\measo_Y$ jointly optimizes the projective measurement and the associated real-valued estimates.
Define
\begin{align*}
\rho_Y & =\Tr_X(\rho_{XY})=(1-p)\rho^0+p\rho^1, \\
D & = (1-p)\rho^0-p\rho^1.
\end{align*}
Then the mean-squared error associated with $\measo_Y$ is
\begin{align}
    \mse(\rho_{XY},\measo_Y)
    &\coloneqq \Tr\big(\rho_{XY}(\obs_X-\obs_Y)^2\big) \nonumber \\
    &=1-2\Tr(D\measo_Y)+\Tr(\rho_Y\measo_Y^2). \label{eq:mmse}
\end{align}

In the short ISIT version of this manuscript~\cite{mandal2025isit},
the statement and proof of the following lemma contained errors,
which have been corrected in the present version.

\begin{lem}\label{mmse-observable}
Consider the CQ state $\rho_{XY}$
and the MSE functional given by~\eqref{eq:mmse}.
If $\rho_Y$ is positive definite, then the MMSE-achieving observable $\measo_Y^*$ is the unique Hermitian solution of the Sylvester equation
\begin{align}\label{eq:lyapunov-mmse}
    \rho_Y\measo_Y+\measo_Y\rho_Y=2D.
\end{align}
For any $\rho_Y \in \dop$ satisfying $\rho_Y=\sum_a \lambda_a \ketbra{e_a}{e_a}$, we have
\begin{align}\label{eq:mmse-solution}
    \bra{e_a}\measo_Y^*\ket{e_b}
    = \begin{cases} 
    \frac{2\bra{e_a}D\ket{e_b}}{\lambda_a+\lambda_b} & \text{if } \lambda_a+\lambda_b>0 \\ 0 & \text{otherwise.} \end{cases}
\end{align}
The minimum value is
\begin{align}
    \mmse(X|Y)_{\rho_{XY}} \msn
    &= \msn
    1 \msn - \msn \Tr(D\measo_Y^*)
    \msn \\
    &= \msn
    1 \msn - \msn \sum_{a,b : \lambda_a+\lambda_b>0} \msn
    \frac{2|\bra{e_a}D\ket{e_b}|^2}{\lambda_a+\lambda_b}. \label{eq:mmse-value}
\end{align}
If \(\rho_Y\) is singular, all entries of an MMSE minimizer having at least one index in \(\supp(\rho_Y)\) are uniquely determined by~\eqref{eq:mmse-solution}.
For entries indexed by two indices orthogonal to the support, the values are arbitrary and the canonical minimizer sets them to zero.
\end{lem}

\begin{proof}
Throughout, we restrict our attention to Hermitian $M_Y$.
Thus, for a Hermitian perturbation $Q_Y$, we have
\begin{align*}
&\frac{d}{d\epsilon}
\mse(\rho_{XY},\measo_Y+\epsilon Q_Y)\bigg|_{\epsilon=0}\\
&
\;\;=
-2\Tr(DQ_Y)+\Tr(\rho_Y\measo_Y Q_Y)+\Tr(\rho_Y Q_Y\measo_Y)\\
&
\;\;=
\Tr\!\left(
(\rho_Y\measo_Y+\measo_Y\rho_Y-2D)Q_Y
\right),
\end{align*}
where the last step uses cyclicity of trace.
Since this must vanish for all Hermitian $Q_Y$, the $\measo_Y$ stationary points must satisfy~\eqref{eq:lyapunov-mmse}.
With $\rho_{XY}$ fixed, $\mse(\rho_{XY},\measo_Y)$ is convex on $\measo_Y \in \hop^d$ because
\begin{multline*}
\mse(\rho_{XY},\measo_Y+Q_Y)-\mse(\rho_{XY},\measo_Y)  \\
= \Tr((\rho_Y\measo_Y+\measo_Y\rho_Y-2D)Q_Y)
+
\Tr(\rho_Y Q_Y^2)
\end{multline*}
and \(\Tr(\rho_Y Q_Y^2)\ge0\).
Thus, all stationary points have the same minimum value. Moreover, since $\mse(\rho_{XY},\measo_Y)$ is strictly convex in the projection of $\measo_Y$ onto the support of $\rho_Y$, the stationary point uniquely determines the minimizer on the support of $\rho_Y$.

Let \(P\) be the support projector of \(\rho_Y\).
Since \(\rho_Y=(1-p)\rho^0+p\rho^1\) is a positive combination of PSD operators, every vector in \(\ker\rho_Y\) lies in the kernel of each conditional state with positive prior.
Hence, \(PD=DP=D\) and the numerator \(\bra{e_a}D\ket{e_b}\) is zero whenever \(\lambda_a+\lambda_b=0\), so the singular Sylvester formula is well defined on \(\supp(\rho_Y)\).

To determine the solution set,
one can use an eigendecomposition $\rho_Y=\sum_a\lambda_a\ketbra{e_a}{e_a}$ and project~\eqref{eq:lyapunov-mmse} via $\bra{e_a}\, \cdot \, \ket{e_b}$ in order to get the equations in~\eqref{eq:mmse-solution}.
It is easy to verify that these equations uniquely determine the operator $\measo_Y^*$ on the support of $\rho_Y$.
Finally, multiplying \eqref{eq:lyapunov-mmse} by $\measo_Y^*$ and taking the trace gives
\[
\Tr(\rho_Y(\measo_Y^*)^2)=\Tr(D\measo_Y^*).
\]
Substituting this into~\eqref{eq:mmse} gives~\eqref{eq:mmse-value}.
\end{proof}

Lemma~\ref{mmse-observable} characterizes the MMSE-achieving observable for a general binary CQ state.
For the BSCQ channels used in the coding problem, the input is uniform and the two output states are related by the channel symmetry.
In this case, the optimizer may be chosen to be antisymmetric under the symmetry operator.
This choice allows the MMSE to be evaluated by conditioning on the input value $0$.
An equivalent expression in terms of the symmetric logarithmic derivative (SLD) quantum $\chi^2$ divergence is given in Appendix~\ref{property mmse}.

\subsection{MMSE Relation for Symmetric BSCQ Channels}

\begin{lem}\label{pm-symmetry}
Consider a BSCQ channel with outputs $\rho^{0}$ and $\rho^{1}$, where $\rho^{1}=U\rho^{0}U^\dagger$ with $U^2=\mI$, and a uniform random input $X$.
Let $\measo_Y^*$ be an MMSE-achieving observable.
Then
\begin{align}\label{eq:allzero-mmse}
    \mmse(X|Y)_{\rho_{XY}}
    =
    \Tr\!\left(\rho^0(\mI_Y-\measo_Y^*)^2\right).
\end{align}
Moreover, the canonical minimizer, which vanishes on $\ker\rho_Y$, satisfies $U\measo_Y^*U^\dagger=-\measo_Y^*$.
\end{lem}

\begin{proof}
For $p=1/2$, define
\[
\rho_Y=\frac{\rho^0+\rho^1}{2},
\qquad
D=\frac{\rho^0-\rho^1}{2}.
\]
The channel symmetry implies
\[
U\rho_YU^\dagger=\rho_Y,
\qquad
UDU^\dagger=-D.
\]
Note that, if \(P\) is the support projector for \(\rho_Y\), then \(UPU^\dagger=P\) and \(U\) preserves \(\supp(\rho_Y)\).

Since $\measo_Y^*$ satisfies
\[
\rho_Y\measo_Y^*+\measo_Y^*\rho_Y=2D,
\]
applying the mapping $\measo \mapsto U \measo U^\dagger$ to both sides shows that
\[
\rho_Y(-U\measo_Y^*U^\dagger)+(-U\measo_Y^*U^\dagger)\rho_Y=2D.
\]
It follows that $-U\measo_Y^*U^\dagger$ solves the same equation.
Since the solution of the Sylvester equation is unique when restricted to the support of $\rho_Y$ and $UPU^\dagger=P$, the projection of $\measo_Y^*$ onto the support of $\rho_Y$ is also antisymmetric under the mapping.
Therefore, we assume without loss of generality that
\[
U\measo_Y^*U^\dagger=-\measo_Y^*.
\]
It follows that
\[
\Tr(\rho_Y\measo_Y^*)=0,
\qquad
\Tr(D(\measo_Y^*)^2)=0,
\]
because $\rho_Y$ is invariant under $U$ whereas $\measo_Y^*$ and $D$ are antisymmetric. 

Since $\rho^0=\rho_Y+D$, we have
\begin{align*}
&\Tr(\rho^0(\mI-\measo_Y^*)^2)\\
&
=
1-2\Tr(\rho^0\measo_Y^*)+\Tr(\rho^0(\measo_Y^*)^2)\\
&
=
1-2\Tr(D\measo_Y^*)+\Tr(\rho_Y(\measo_Y^*)^2).
\end{align*}
This is exactly the MSE attained by $\measo_Y^*$.
\end{proof}

The recursion will be stated for $\mmse$, but the final coding theorem is stated in terms of bit-error probability.  
Therefore, we need inequalities that relate the MMSE of the optimal observable to the Helstrom error probability.
The next lemma gives both directions needed later.
One direction converts MMSE decay into bit-error decay, and the other converts an initial entropy bound into an initial MMSE bound.

\subsection{Relation Between MMSE and Helstrom Error}
\begin{lem}\label{mmse-p-error}
    Consider a BSCQ with outputs $\rho^{0}$ and $\rho^{1}$, where $\rho^{1}=U\rho^{0}U^\dagger$ with $U^2=\mI$, and a uniform random input $X$ (i.e. $p=1/2$).
    Then, the MMSE $\mmse(X|Y)_{\rho_{XY}}$ satisfies
\begin{align*}
    \mmse(X|Y)_{\rho_{XY}} &\ \geq 2P_{e}\left(\rho^0,\rho^1,\frac{1}{2}\right)\\
    \mmse(X|Y)_{\rho_{XY}} &\ \leq 4P_{e}\left(\rho^0,\rho^1,\frac{1}{2}\right)\left(1-P_{e}\left(\rho^0,\rho^1,\frac{1}{2}\right)\right).
\end{align*}

\end{lem}

\begin{proof}
Let
\[
\measo_Y^*=\sum_j m_j P_j
\]
be the spectral decomposition of the MMSE-achieving observable, where the $P_j$ are orthogonal projectors. Define
\[
p_j^x \coloneqq \Tr(\rho^x P_j), \qquad
p_j \coloneqq \frac{p_j^0+p_j^1}{2}
\]
and let $J$ be the set of indices $j$ such that $p_j>0$.
For this fixed projective measurement, the MSE becomes
\[
\mse(\rho_{XY},\measo_Y^*)
=
\sum_{j\in J} p_j\left(1-2m_j\frac{p_j^0-p_j^1}{p_j^0+p_j^1}+m_j^2\right).
\]
Hence, for fixed $j\in J$, the minimizing coefficient is
\[
m_j=\frac{p_j^0-p_j^1}{p_j^0+p_j^1}
\]
and we define $\eta_j \coloneqq (1-m_j)/2$.
Thus, we find
\[
\mmse(X|Y)_{\rho_{XY}}
=
\sum_{j \in J} 4p_j\eta_j(1-\eta_j).
\]
For the same measurement $\{P_j\}$, the optimal classical decision rule has error probability
\[
P_e^{\{P_j\}}
=
\sum_{j\in J} p_j \min(\eta_j,1-\eta_j).
\]
Since $4t(1-t)\ge 2t$ for every $t\in[0,1/2]$, we observe that
\[
\mmse(X|Y)_{\rho_{XY}}
\ge 2 P_e^{\{P_j\}}
\ge 2 P_e\!\left(\rho^0,\rho^1,\tfrac{1}{2}\right)
\]
because the Helstrom error is the minimum possible.

For the upper bound, let $\Pi_0$ be a Helstrom projector and define
\[
\widetilde{\measo}_Y=\alpha(2\Pi_0-\mI_Y).
\]
Since $(2\Pi_0-\mI_Y)^2=\mI_Y$, the MSE of this trial observable is
\begin{align*}
\mse(\rho_{XY},\widetilde{\measo}_Y)
&=
1-2\alpha\,\Tr\!\left(D(2\Pi_0-\mI_Y)\right)+\alpha^2.
\end{align*}
By Helstrom's formula,
\[
\Tr\!\left(D(2\Pi_0-\mI_Y)\right)
=
\frac{1}{2}\|\rho^0-\rho^1\|_1
=
1-2P_e\!\left(\rho^0,\rho^1,\tfrac{1}{2}\right).
\]
Choosing $\alpha=1-2P_e(\rho^0,\rho^1,\tfrac{1}{2})$ yields
\[
\mse(\rho_{XY},\widetilde{\measo}_Y)
=
4P_e\!\left(\rho^0,\rho^1,\tfrac{1}{2}\right)
\left(1-P_e\!\left(\rho^0,\rho^1,\tfrac{1}{2}\right)\right).
\]
Since $\measo_Y^*$ minimizes the MSE, the same upper bound holds for $\mmse(X|Y)_{\rho_{XY}}$.
\end{proof}

\section{Reed--Muller Codes over BSCQ Channels}\label{sec:rm-cq-decoding}

In this section, we apply the binary CQ setup to the extrinsic channel induced by the $\RM$ code.
A uniformly random codeword $\bX\in\cC$ is transmitted through independent uses of the BSCQ channel, and the receiver estimates $X_0$ using only the quantum systems $Y_{\sim0}$.
The two conditional output states of this induced binary channel are obtained by averaging over the codewords with $x_0=0$ and $x_0=1$.

Since $\RM$ codes are transitive and we consider transmission over memoryless BSCQ channels, any performance metric for a single bit must take the same value for all bits.
Thus, without loss of generality, we focus on bit 0 for our analysis.
Recent analyses of $\RM$ codes over classical BMS channels are based on finding an appropriate boolean function that characterizes the decodability of a single bit from the extrinsic observation~\cite{Reeves-it23,Reeves-isit23,Abbe-focs23,Pfister-arxiv25a}.
This section will introduce the necessary concepts to set up $\RM$ code decoding for a single bit from the extrinsic observation over BSCQ channels.

Consider the $\RM$ code $\cC = \RM (r,m)$ of length $N = 2^{m}$.
As discussed in Proposition~\ref{prop:rm_prop}(c), one can partition the $n=N-1$ non-zero indices into disjoint sets $A,B,C,D\subseteq [N]\setminus \{0\}$ such that the code projections $\cC' = \cC|_{\{0\}\cup A\cup B}$ and $\cC'' = \cC|_{\{0\} \cup A \cup C}$ both equal $\RM(r,m-1)$.

Let $\bX$ denote a uniform random codeword from $\cC$ and $\bY$ denote the $N$ quantum systems that define the output when transmitting $\bX$ through independent BSCQ channels.
We use $Y_{\sim 0}$ to denote the quantum systems used for extrinsic estimation of $X_0$.
Then, the associated CQ state is given by
\begin{align*}
    \rho_{\bX Y_{\sim 0}}=\frac{1}{|\cC|}\sum_{\bm{x}\in \cC} \ketbra{\bm{x}}{\bm{x}}_{\bX}\otimes \rho^{x_{\sim 0}}_{Y_{\sim 0}},
\end{align*}
where $\rho^{x_{\sim 0}}_{Y_{\sim 0}}=\rho^{x_1}_{Y_{1}}\otimes \dots \otimes \rho^{x_{n}}_{Y_{n}}$.
For codes $\cC'$, $\cC''$ the extrinsic CQ states $\rho_{X_{\{0\}\cup A\cup B}Y_{ A\cup B}}$ and $\rho_{X_{\{0\}\cup A\cup C}Y_{ A\cup C}}$ are defined analogously.

To obtain the error probability of decoding bit-0 of $\cC$ via extrinsic observation of the quantum system $Y_{\sim 0}$, we compute
\begin{align}
    \rho_{X_0Y_{\sim 0}} &\ = \Tr_{X_{ \sim 0}}\left( \rho_{\bX Y_{\sim 0}}\right)
     = \frac{1}{2}\sum_{z\in\{0,1\}}\ketbra{z}{z}
    _{X_{0}}\otimes \overline{\rho}^{z}_{Y_{\sim 0}},
\end{align}
where 
\vspace*{-3.5mm}
\begin{align*}
  \overline{\rho}^{z}_{Y_{\sim 0}}=\frac{2}{|\cC|}\sum_{\bm{x}\in \cC,x_0=z}\rho^{x_{\sim 0}}_{Y_{\sim 0}}.
\end{align*}
Thus, the extrinsic bit-error probability for $\cC$ is given by $ P_{b}(\cC) =P_{e}\left( \overline{\rho}^{0}_{Y_{\sim 0}},  \overline{\rho}^{1}_{Y_{\sim 0}},\frac{1}{2}\right).$
For codes $\cC'$, $\cC''$ the extrinsic CQ states $\overline{\rho}^{x}_{Y_{ A\cup B}},\overline{\rho}^{x}_{Y_{ A\cup C}}$ and bit-error probabilities $P_{b}(\cC'),P_{b}(\cC'')$ are defined analogously.

Let \(\Pi_+,\Pi_-,\Pi_\circ\) be the positive, negative, and zero spectral projectors of \(\overline\rho^0_{Y_{\sim0}}-\overline\rho^1_{Y_{\sim0}}\) so that the optimal two-outcome POVM is defined by \(E_0=\Pi_+ +\Pi_\circ/2\), \(E_1=\Pi_- +\Pi_\circ/2\).
The all-ones symmetry exchanges \(E_0\) and \(E_1\), whereas every zero-coset symmetry fixes each effect.
This implies that
\begin{align*}
    P_{b}(\cC)
    &= \frac{1}{2}\Tr\left(E_1 \overline{\rho}^{0}_{Y_{\sim 0}}\right) + \frac{1}{2}\Tr\left(E_0 \overline{\rho}^{1}_{Y_{\sim 0}}\right)\\
     &= \Tr(E_1\overline\rho^0) \\
     &=\Tr(E_1(\rho^0)^{\otimes n}).
\end{align*}
The second step holds because the all-ones unitary interchanges the conditional states and satisfies $U^{\otimes n} E_0 U^{\otimes n} = E_1$ and the third step follows because the strengthened punctured form of Lemma~\ref{lem:helstrom-code-symmetry} applies to \(E_1\), which is invariant under conjugation by codewords in the zero coset.
Thus, the extrinsic bit-error probability may be evaluated under the all-zero transmitted codeword.
Also, the balanced POVM is defined here only to make the $0/1$ exchange symmetry exact.
Any projective allocation of the zero eigenspace has the same average Helstrom error and may be used instead when a sequence of projectors is required.

Thus, one can analyze the extrinsic bit-error probability by considering only the extrinsic observation from the all-zero codeword  $(\rho^0)^{\otimes n}$.
Since the all-zero word is one term in the \(x_0=0\) coset average, \(\supp((\rho^0)^{\otimes n})\subseteq\supp(\overline{\rho}^{0}_{Y_{\sim0}})\).
Also \(\supp(\overline{\rho}^{0}_{Y_{\sim0}})\subseteq\supp((\overline{\rho}^{0}_{Y_{\sim0}}+\overline{\rho}^{1}_{Y_{\sim0}})/2)\).
Therefore, in the rank-deficient case, the MMSE observable for the extrinsic channel is uniquely determined on every vector that can contribute to the all-zero expectations used below.

Let \(W_{\mathrm{extr}}^\cC:\{0,1\}\to\dop\), \(x_0\mapsto\overline\rho^{x_0}_{Y_{\sim0}}\), denote the binary CQ channel induced by \(\rho_{X_0Y_{\sim0}}\).  This is the BSCQ channel for decoding bit \(0\) of \(\cC\) from the extrinsic quantum system \(Y_{\sim0}\).
In particular, addition of the all-ones codeword interchanges the cosets with $x_0=0$ and $x_0=1$.
Thus, the tensor product \(U^{\otimes N-1}\) of the symmetry unitary for $W$ maps \(\overline\rho^0_{Y_{\sim0}}\) to \(\overline\rho^1_{Y_{\sim0}}\).




\section{Extrinsic Information Transfer Functions}\label{sec:exit-functions}


An EXIT function describes how the information about one coordinate changes as the observations of the other coordinates are gradually erased.
This provides a way to relate the uncertainty of the extrinsic observation $Y_{\sim 0}$ to the rate of the code.
In particular, we introduce an erasure interpolation between the original BSCQ channel and a channel whose output is always erased.
Along this interpolation, the change in the conditional entropy of the transmitted codeword can be decomposed into single-coordinate extrinsic information terms.
For a transitive code, these terms are the same for every coordinate, which gives an EXIT-area identity for a single coordinate.
We use this identity to convert a rate gap $C-R$ into a nontrivial bound on the extrinsic uncertainty of one coordinate.

For the BSCQ channel $W$,
let $W_t$ denote the BSCQ channel in which the quantum output of $W$ is erased with probability $t$.
For the channel input vector $\bx = (x_0,\ldots,x_{N-1})$, the corresponding quantum output on quantum systems $\bY = (Y_0,Y_1,\ldots,Y_{N-1})$ is defined by the density matrix
\[ W_t (\bx) = W_t (x_0) \otimes W_t (x_1) \otimes \cdots \otimes W_t (x_{N-1}). \]
We will sometimes drop the $t$ subscript when $t=0$ or alternatively write $\bY (t)$ to emphasize the dependence on $t$.

Let $\cC \subseteq \cX^N$ be a length-$N$ code for this CQ channel.
For a random codeword $\bX \in \cC$, drawn according to $P_{\bX} (\bx)$, the resulting density matrix is
\[ \rho_{\bX\!\bY (t)} = \sum_{\bx \in \cC} P_{\bX} (\bx) \left( \ket{\bx}\bra{\bx} \otimes W_t (\bx) \right). \]
We follow the convention of~\cite{Wilde-2013} and denote the mutual information by
\[ I\big(\bX;\bY(t) \big)_{\rho_{\bX\!\bY (t)}} .\]

The erasure parameter $t$ provides an interpolation path from the fully observed case to complete erasure.
Differentiating the conditional entropy along this path decomposes the total information into single-coordinate extrinsic contributions.
For transitive codes, these contributions are identical across coordinates, so the global mutual information per bit is equal to the area under the bitwise extrinsic information curve.

\begin{lem}\label{exit-cq}
For a binary code $\cC$ with codewords chosen from the distribution $P_{\bX}$ on a binary-input CQ channel, we have
\begin{align*}
\frac{1}{N} I(\bX ; \bY)_{\rho_{\bX\!\bY}}
&= \frac{1}{N} \int_0^1 \sum_{i=0}^{N-1} I\big (X_i; Y_i \mid Y_{\sim i}(t) \big)_{\rho^i_{\bX\!\bY (t)}} \, dt,
\end{align*}
where $\rho^{i}_{\bX\!\bY (t)}$ is the density matrix for the system where $Y_i$ does not experience erasures.    
If the automorphism group of $\cC$ is transitive (i.e., $\cC$ has transitive symmetry)
and the codeword distribution is uniform, then this reduces to\vspace{-0.5mm}
\begin{align*}
\frac{1}{N} I(\bX ; \bY)_{\rho_{\bX\!\bY}}
&= \int_0^1 I\big (X_0; Y_0 \mid Y_{\sim 0}(t) \big)_{\rho^0_{\bX\!\bY (t)}} \, dt. 
\end{align*}

\end{lem}

\begin{proof}
First, we note that
\begin{align*}
I(\bX ; \bY)_{\rho_{\bX\!\bY}}
&= H(\bX) - H(\bX | \bY)_{\rho_{\bX\!\bY}} \\
&= H(\bX | \bY(1))_{\rho_{\bX\!\bY(1)}} - H(\bX | \bY(0))_{\rho_{\bX\!\bY(0)}} \\
&= \int_0^1 \frac{d}{dt} H(\bX | \bY(t))_{\rho_{\bX\!\bY(t)}} \, dt. 
\end{align*}
Next, we assume that the erasure rate for each system is different (e.g., the erasure rate for $Y_i$ is $t_i$) and use the law of the total derivative to write
\begin{align*}
&\ \frac{d}{dt} H(\bX | \bY(t))_{\rho_{\bX\!\bY(\bt)}} \\
&\ = \sum_{i=0}^{N-1} \frac{d}{dt_i} H(\bX | \bY(\bt))_{\rho_{\bX\!\bY(\bt)}} \bigg|_{t_0 = t_1 = \cdots = t_{N-1} = t},
\end{align*}
where $\bt = (t_0,t_1,\ldots,t_{N-1})$ is the vector of erasure rates.
To simplify this expression, we use the chain rule of entropy and the definition of an erasure channel to write
\begin{align*}
&\ H(\bX | \bY(\bt))_{\rho_{\bX\!\bY(\bt)}}\\
&\ = H(X_i | \bY(\bt))_{\rho_{\bX\!\bY(\bt)}}
+H(X_{\sim i} | \bY(\bt),X_i)_{\rho_{\bX\!\bY(\bt)}} \\
&\ = (1-t_i) H(X_i | Y_{\sim i} (t_{\sim i}), Y_i)_{\rho^i_{\bX\!\bY(\bt)}} \\
&\ \;\, +
t_i H(X_i | Y_{\sim i} (t_{\sim i}))_{\rho^i_{\bX\!\bY(\bt)}} \!\! + H(X_{\sim i}\mid Y_{\sim i}(t_{\sim i}),X_i)_{\rho_{\bX\!\bY(\bt)}},
\end{align*}
where $\rho^{i}_{\bX\!\bY (\bt)} = \rho_{\bX\!\bY (\bt)}|_{t_i =0}$ is the density matrix of the same system except $Y_i$ does not experience erasures.
Thus, we see
\begin{align*}
&\ \frac{d}{dt_i} H(\bX | \bY(\bt))_{\rho_{\bX\!\bY(\bt)}}\\
&= H(X_i | Y_{\sim i} (t_{\sim i}))_{\rho^i_{\bX\!\bY(\bt)}} -  H(X_i | Y_{\sim i} (t_{\sim i}), Y_i)_{\rho^i_{\bX\!\bY(\bt)}} \\
&= I(X_i;Y_i | Y_{\sim i}(t_{\sim i}))_{\rho^i_{\bX\!\bY(\bt)}}.
\end{align*}
Thus, we observe that
\begin{align*}
&\ \frac{d}{dt} H(\bX | \bY(t))_{\rho_{\bX\!\bY(\bt)}}\\
&= \sum_{i=0}^{N-1} \frac{d}{dt_i} H(\bX | \bY(\bt))_{\rho_{\bX\!\bY(\bt)}}  \bigg|_{t_0 = t_1 = \cdots = t_{N-1} = t} \\
&= \sum_{i=0}^{N-1} I(X_i;Y_i | Y_{\sim i}(t_{\sim i}))_{\rho^i_{\bX\!\bY(\bt)}}.
\end{align*}
Since the channel uses have full permutation symmetry when $t_i = t$ for all $i \in [N]$, if the code has transitive symmetry and the codeword distribution is uniform, then the summand does not depend on $i$ and one gets the stated simplification.
\end{proof}

Lemma~\ref{exit-cq} expresses the normalized information in terms of the extrinsic information of a single coordinate.
The next step is to lower bound the integrand using the channel capacity and the information already available from the other coordinates.
This converts the rate condition $R<C$ into a pointwise upper bound on the extrinsic entropy $H(X_0|Y_{\sim0})_{\rho_{\bX\!\bY}}$.

\begin{lem}\label{rate-relation-cq}
Let $\cC$ be a transitive binary linear $[N,K]$ code with $R=K/N>0$ and let $\bX$ be uniform on $\cC$.  Then,
\[ H(X_0 | Y_{\sim 0} )_{\rho_{\bX\!\bY}} \leq 1 - (C-R). \]
\end{lem}

\begin{proof}
We start by writing
\begin{align*}
C &= I(X_0; Y_0)_{\rho_{X_0 \! Y_0}} \\
&= H(Y_0)_{\rho_{X_0 \! Y_0}} - H(Y_0|X_0)_{\rho_{X_0 \! Y_0}} \\
&= H(Y_0)_{\rho^0_{\bX\!\bY(t)}} - H(Y_0|X_0,Y_{\sim 0} (t) )_{\rho^0_{\bX\!\bY (t)}} \\
&= I(Y_0; X_0, Y_{\sim 0} (t) )_{\rho^0_{\bX\!\bY(t)}} \\
&= I(Y_0; Y_{\sim 0}(t) )_{\rho^0_{\bX\!\bY(t)}} + I(Y_0;X_0 | Y_{\sim 0} (t))_{\rho^0_{\bX\!\bY(t)}} \\
&\leq I(X_0; Y_{\sim 0} (t) )_{\rho^0_{\bX\!\bY(t)}} + I(Y_0;X_0 | Y_{\sim 0}(t) )_{\rho^0_{\bX\!\bY(t)}} \\
&\leq I(X_0; Y_{\sim 0} )_{\rho_{\bX\!\bY}} + I(Y_0;X_0 | Y_{\sim 0}(t) )_{\rho^0_{\bX\!\bY(t)}},
\end{align*} 
where the first inequality follows from the Markov structure \(Y_0 - X_0 - Y_{\sim0}(t)\) and, conditional on $X_0$, this implies $Y_0$ is independent of $Y_{\sim0}(t)$ because the physical channel is memoryless.
The second inequality holds because $Y_{\sim 0}(t)$ is a degraded version of $Y_{\sim 0}$ and $\rho^0_{\bX\!\bY} = \rho^0_{\bX\!\bY(0)} = \rho_{\bX\!\bY}$.
Now, we can use the EXIT area theorem to write
\begin{align*}
\frac{1}{N} H(\bX)
&\geq \frac{1}{N} I(\bX;\bY) \\
&= \int_0^1 I\big (X_0; Y_0 \mid Y_{\sim 0}(t) \big)_{\rho^0_{\bX\!\bY (t)}} \, dt \\
&\geq \int_0^1 \big( C - I(X_0; Y_{\sim 0} )_{\rho_{\bX\!\bY}} \big) dt \\
&= C - I(X_0; Y_{\sim 0} )_{\rho_{\bX\!\bY}}
\end{align*}
Since the code is transitive and \(R>0\), every \(X_i\) is uniform on $\{0,1\}$ (i.e., no code bits are identically zero).
This shows that $H(X_0)=1$ and implies the desired conclusion $R \geq C - 1 + H(X_0|Y_{\sim 0})_{\rho_{\bX\!\bY}}$.
\end{proof}

The entropy bound above identifies when the extrinsic channel for one coordinate is already informative.

\section{Orthogonal Decomposition of Observables}\label{sec:orthogonal-decomposition}

Decoder error can be naturally expressed in terms of quantum observables.
Thus, we introduce a Fourier-type decomposition for quantum observables with respect to the product background state.
This decomposition enables the proof of a correlation inequality for two observables that depend on distinct but overlapping subsets of quantum systems.

To show a vanishing bit-error probability for $\RM$ codes on BSCQ channels, we rely on a suitable choice of inner product to decompose quantum observables. For our analysis, we adopt the complex inner product established from  Gelfand–Naimark–Segal (GNS) construction~\cite{arveson2012invitation}. 
For any full-rank $\rho \in \dop^d$ and $\obsf,\obsg \in \mathbb{C}^{d\times d}$, consider the GNS inner product on $\mathbb{C}^{d\times d}$ defined by
\[ \inner{\obsf}{\obsg}_\rho \triangleq \Tr(\obsg^{\dagger} \rho \obsf). \]

For the Hilbert space $\mathbb{C}^{d\times d}$ over $\mathbb{C}$, there exists an orthonormal basis $\{ \basis{0}, \basis{1}, \ldots, \basis{d^2 -1} \}$ with $\basis{0} = \mI$.
For example, one can apply Gram-Schmidt to any basis whose first element is $\mI$.
Although the GNS inner product can be complex, the $\basis{0}$-coefficient of a Hermitian $\obsf$ is real because $ \inner{\obsf}{\basis{0}}_\rho = \Tr(\rho \obsf)$ and \[\inner{ \obsf }{ \basis{0} }_\rho^\dagger = \inner{ \basis{0} }{ \obsf }_\rho = \Tr(\obsf^\dagger \rho) = \Tr(\rho \obsf).\]
Since this inner product is positive definite only if $\rho$ is full-rank, a modified version is defined in Lemma~\ref{lem:gns-singular} for any $\rho$.

For $n$ qudits, this extends to $\mathbb{C}^{d^n\times d^n}$ with the inner product \vspace{-2mm}
\[ \innerwgn{\obsf}{\obsg} \coloneqq \Tr(\obsg^{\dagger} \measgn \obsf) \]
and the single-system basis extends to an orthonormal basis $\{\basis{\bs}\}_{\bs \in [d^2]^n}$ for $\mathbb{C}^{d^n\times d^n}$.
Specifically, for $\bs \in [d^2]^n$, we have \[ \basis{\bs} \triangleq \basis{s_0} \otimes \cdots \otimes \basis{s_{n-1}}. \]
Using this basis, we can write any observable $\obsf \in \mathbb{C}^{d^n\times d^n}$ as
\begin{align*}
\obsf = \sum_{\bs \in [d^2]^n} \F{\obsf}_{\bs} \basis{\bs}, \qquad \F{\obsf}_{\bs} \coloneqq \innerwgn{\obsf}{\basis{\bs}}.
\end{align*}

The study of quantum boolean functions is closely related to the orthogonal decomposition of observables.
Until now, most work in this area is restricted to the uniform background measure where $\rho=\mI$ \cite{montanaro2008quantum}.
For classical boolean functions, there is a well-developed theory beyond this (e.g., $p$-biased measures).
In this context, Parseval's theorem implies that
\[ \normwgn{\obsf}^2 = \Tr(\obsf^{\dagger} \rho^{\otimes n} \obsf) = \sum_{\bs \in [d^2]^n} |\F{\obsf}_{\bs}|^2 . \]
Thus, we find that
\[ \text{Var}(\obsf)_{\measgn} = \normwgn{\obsf}^2 - \left|\innerwgn{\obsf}{\mI}\mkern-1mu \right|^2 = \sum_{\bs \in [d^2]^n : \bs \neq \bm{0}} |\F{\obsf}_{\bs}|^2 . \]
The above description shows that \(\innerwgn{\obsf}{\mI}=\F{\obsf}_{\mathbf{0}}\) is real for Hermitian \(\obsf\).

We say that an observable $F$ on $n$ systems is \emph{supported within} the subset $A \subseteq [n]$ if we can write $\obsf = \obsf_A \otimes \, \mI_{\sim A}$ where $\obsf_A \in \hop^{d^{|A|}}$ and the tensor product is taken to preserve the original order of the quantum systems.
Its minimal support is the smallest such $A$.
For an arbitrary subset $A\subseteq [n]$, we define $\cS_{A}$ to be the collection of non-trivial strings $\bs \in [d^2]^n$ supported within $A$,  \[\cS_{A}\triangleq \{\bs \in [d^2]^n \,|\, s_i=0, \forall i\notin A\}-\{\mathbf{0}\}. \] 
We also note that $\basis{\bs} = \basis{\bs_A} \otimes \, \basis{\bs_{\sim A}}$, where the tensor product is taken to preserve the original order of the quantum systems.

Now, we will see the benefit of choosing our single-system basis with $\Omega_{0} = \mI$.
Based on this choice, if $\obsf$ is supported within a non-empty set $A$ and $\bs \neq \mathbf{0}$, then it follows that
\begin{align*}
\F{\obsf}_{\bs}
&= \Tr\left( (\basis{\bs_A} \otimes \basis{\bs_{\sim A}})^{\dagger} \rho^{\otimes n}  (\obsf_A \otimes \mI_{\sim A})\right) \\
&= \Tr\left(\basis{\bs_A}^{\dagger} \, \rho^{\otimes |A|} \, \obsf_A   \right)
\Tr\left( \basis{\bs_{\sim A}}^{\dagger} \, \rho^{\otimes |\sim A|} \, \mI_{\sim A} \right) \\
&= \begin{cases} \Tr\left( \basis{\bs_A}^{\dagger} \, \rho^{\otimes |A|} \,  \obsf_A \right) & \text{if $\bs \in \cS_A$} \\ 0 & \text{otherwise} \end{cases}
\end{align*}
because $\Tr \big( \basis{\bs_{\sim A}}^{\dagger} \, \rho^{\otimes |\sim A|} \, \mI_{\sim A} \big) = 0$ if $\bs \notin \cS_A \cup \{\mathbf{0}\}$.

The preceding GNS construction assumes that $\rho$ is full rank.
But, when $\rho$ is rank deficient, the inner product is semidefinite on the full operator space, but may be positive definite on a subspace.
In general, all GNS inner products depend only on the left-supported representatives $PF$ and $PG$, where $P$ is the projection onto $\supp(\rho)$.
The next lemma shows that restricting to these representatives gives a genuine Hilbert space and preserves all elements necessary for the correlation inequality.

\begin{lem}\label{lem:gns-singular}
Let $\rho\in \dop^d$ and let $P$ be the orthogonal projector onto $\supp(\rho)$.  For $F,G\in\mathbb C^{d\times d}$, we have
\begin{align*}
    \langle F,G\rangle_{\rho}
    =
    \langle PF,PG\rangle_{\rho}.
\end{align*}
Moreover, $\langle\cdot,\cdot\rangle_{\rho}$ is positive definite on the subspace spanned by left-support GNS representatives,
\begin{align*}
    \cD = \{F\in\mathbb C^{d\times d}:PF=F\}.
\end{align*}
Restricting observables to this subspace gives a well-defined Hilbert space of dimension $d\operatorname{rank}(\rho)$.
Such a space always contains an orthonormal basis of the form
\begin{align*}
    \{\Omega_0,\Omega_1,\ldots,\Omega_{d\operatorname{rank}(\rho)-1}\},
    \qquad
    \Omega_0=P.
\end{align*}
For $n$ qudits, the tensor-product basis $\{\Omega_{\bs}\}_{\bs\in[d\operatorname{rank}(\rho)]^n}$ is orthonormal with respect to $\langle\cdot,\cdot\rangle_{\rho^{\otimes n}}$, and every $F\in\mathbb C^{d^n\times d^n}$ satisfies
\begin{align*}
    P^{\otimes n}F
    =
    \sum_{\bs\in[d\operatorname{rank}(\rho)]^n}\widehat F_{\bs}\Omega_{\bs},
    \qquad
    \widehat F_{\bs}
    :=
    \langle F,\Omega_{\bs}\rangle_{\rho^{\otimes n}},
\end{align*}
with Parseval identity
\begin{align*}
    \normwgn{F}^{2}
    =
    \sum_{\bs\in[d\operatorname{rank}(\rho)]^n}|\widehat F_{\bs}|^2 .
\end{align*}
\end{lem}

\begin{proof}
Since $P$ is the support projection of $\rho$, we have
\begin{align*}
    P\rho=\rho P=\rho.
\end{align*}
Thus, we have
\begin{align*}
    \langle F,G\rangle_{\rho}
    &=
    \Tr(G^\dagger \rho F)\\
    &=
    \Tr((PG)^\dagger \rho (PF))\\
    &=
    \langle PF,PG\rangle_{\rho}.
\end{align*}
If $PF=F$ and $\langle F,F\rangle_{\rho}=0$, then
\begin{align*}
    \Tr(F^\dagger\rho F)=0.
\end{align*}
Since $\rho$ is positive definite on $\supp(\rho)$, this implies $F=0$.
Hence, the inner product is positive definite on $\cD$.  Since $\cD$ is finite-dimensional, it is a Hilbert space.

The tensor-product statement follows from
\begin{align*}
    \langle \Omega_{\bs},\Omega_{\bt}\rangle_{\rho^{\otimes n}}
    =
    \prod_{i=0}^{n-1}
    \langle \Omega_{s_i},\Omega_{t_i}\rangle_{\rho}
    =
    \delta_{\bs,\bt}.
\end{align*}
Therefore the expansion of $P^{\otimes n}F$ and Parseval's identity follow from ordinary orthonormal-basis expansion.
\end{proof}

\begin{rem}
For brevity, the Fourier results later are based on the full-rank case where the Fourier indices live in $[d^2]^n$ and $\Omega_0=\mI$.
But, all later Fourier arguments apply to rank-deficient $\rho$ with minor changes.

In particular, we let $P$ be the projection onto $\supp(\rho)$, define $q=d\operatorname{rank}(\rho)$, and use the tensor basis for $\cD$ indexed by $[q]^n$ with $\Omega_0=P$.
If $F=F_A\otimes \mI_{\sim A}$, then $P^{\otimes n}F=(P^{\otimes|A|}F_A)\otimes P^{\otimes|\sim A|}$, so its coefficients vanish outside $\cS_A\cup\{\mathbf0\}$.
Moreover, $\swap_\pi P^{\otimes n}\swap_\pi^\dagger=P^{\otimes n}$ and $\swap_\pi\Omega_{\bs}\swap_\pi^\dagger=\Omega_{\pi\bs}$.
Therefore, Lemma~\ref{observable-symmetry} and the proof of Lemma~\ref{cq-transitiveQ} apply verbatim after replacing $[d^2]^n$ by $[q]^n$ and replacing $\mI$ by $P$ in tensor products.
The last point, with $\rho P = \rho$, implies that the inner product still satisfies $\inner{ \obsf }{ \basis{0} }_\rho = \inner{\obsf}{P}_\rho = \Tr(\rho \obsf)$.
\end{rem}

For a permutation $\pi\in \symn $, let $\swap_{\pi}$ be the unitary swap operator satisfying $\swap_{\pi} \basis{\bs} \swap_{\pi}^\dagger = \basis{\pi \bs}$ for $\bs\in [d^2]^n$.
Then, we define the symmetry group of a matrix $\obsf \in \mathbb{C}^{d^n\times d^n}$ to be
\begin{align*}
    \mathrm{Sym}(\obsf)\coloneqq \{\pi\in \symn \,|\,\swap_{\pi}\,\obsf\, \swap_{\pi}^{\dagger}=\obsf\}.
\end{align*}

The next lemma identifies how permutation symmetry appears in the GNS decomposition.
If an observable is invariant under a permutation of the quantum systems, then its GNS coefficients are unchanged after applying the same permutation to their indices.
This will be used to show that, under a transitive symmetry assumption, certain sums of squared Fourier coefficients are the same for all coordinates in the support.

\begin{lem}\label{observable-symmetry}
    For $\pi\in \mathrm{Sym}(\obsf)$ and $\bm{s}\in [d^{2}]^n$, we have 
    \begin{align*}
        \F{\obsf}_{\pi \bm{s}}=\F{\obsf}_{\bm{s}} 
    \end{align*}
\end{lem}

\begin{proof}
    Notice that $\swap_{\pi}\measgn \swap_{\pi}^{\dagger}=\measgn$ $\forall \pi \in \symn$.
Then expanding $\F{\obsf}_{\pi\bm{s}}$ using GNS decomposition for basis element $\Omega_{\pi \bm{s} }$, observable $\obsf$ and measure $\measgn$ we have
\begin{align*}
    \F{\obsf}_{\pi\bm{s}} &\ =\Tr(\Omega_{\pi \bm{s}}^{\dagger}\measgn \obsf )\\
    &\ =\Tr(\swap_{\pi}\Omega_{\bm{s}}^{\dagger}\swap_{\pi}^{\dagger}\measgn \swap_{\pi}\obsf\swap_{\pi}^{\dagger})\\
    &\ = \Tr(\Omega_{\bm{s}}^{\dagger}\measgn \obsf)\\
    &\ =\F{\obsf}_{\bm{s}} . \qedhere
\end{align*}
\end{proof}

We now prove the correlation bound used in the recursive MMSE proof.  The observables $\obsf$ and $\obsg$ are supported within $A\cup B$ and $A\cup C$, respectively, so the only coordinates that can contribute to both observables are the coordinates in $A$.  If $\obsg$ is obtained from $\obsf$ by a permutation that fixes $A$, then the part of $\innerwgn{\obsf}{\obsg}$ that remains after orthogonality is exactly the contribution from the coefficients indexed by $\cS_A$, together with the constant coefficient.  The transitivity assumption on $\mathrm{Sym}(\obsf_{AB})$ then bounds this contribution by averaging over the coordinates in $A\cup B$.

\begin{rem}\label{rem:extrinsic-relabel}
For the decoding analysis of coordinate $0$, set $n=N-1$ and identify the physical labels $[N]\setminus\{0\}$ with $[n]$ via the bijection $j\mapsto j-1$.
The sets $A,B,C$ in the next lemma are the respective images of the previous $A,B,C$ codebit index sets under this relabeling.
Hence, the RM application requires a relabeling of the $n=N-1$ systems in $Y_{\sim0}$ by $[n]$. 
\end{rem}

\begin{figure}
    \centering

    \definecolor{chartreuse}{rgb}{0.5,1.0,0.0}
    \definecolor{darkorange}{rgb}{1.0,0.55,0.0}

    \scalebox{0.5}{%
    \begin{tikzpicture}[
        every node/.style={font=\small},
        site/.style={circle,fill,inner sep=2.4pt},
        Bdot/.style={site,fill=blue!70!black},
        Adot/.style={site,fill=blue!80!black},
        Cdot/.style={site,fill=blue!70!black},
        ABstyle/.style={
            draw=black,
            line width=1.5pt,
            fill=chartreuse,
            rounded corners=15pt,
            opacity=0.5
        },
        ACstyle/.style={
            draw=black,
            line width=1.5pt,
            fill=darkorange,
            rounded corners=15pt,
            opacity=0.5
        }
    ]

    \def\xb{-4}
    \def\yb{0.7}
    \def\dxb{0.9}
    \coordinate (B1) at (\xb,\yb);
    \coordinate (B2) at (\xb+\dxb,\yb);
    \node at (\xb+2*\dxb,\yb-0.6) {\Large $\boldsymbol{B}$};

    \coordinate (B3) at (\xb+2*\dxb,\yb);
    \coordinate (B4) at (\xb+3*\dxb,\yb);
    \coordinate (B5) at (\xb+4*\dxb,\yb);

    \def\xa{0.8}
    \def\ya{0.7}
    \coordinate (A1) at (\xa,\ya);
    \coordinate (A2) at (\xa+0.8,\ya);
    \node at (\xa+1.2,\ya+0.6) {\Large $\boldsymbol{A}$};
  
    \coordinate (A3) at (\xa+2*0.8,\ya);
    \coordinate (A4) at (\xa+3*0.8,\ya);

    \def\xc{4}
    \def\yc{0.7}
    \def\dxc{0.9}
    \coordinate (C1) at (\xc,\yc);
    \coordinate (C2) at (\xc+\dxc,\yc);
    \coordinate (C3) at (\xc+2*\dxc,\yc);
    \node at (\xc+2.5*\dxc,\yc-0.6) {\Large $\boldsymbol{C}$};
    \coordinate (C4) at (\xc+3*\dxc,\yc);
    \coordinate (C5) at (\xc+4*\dxc,\yc);
    \coordinate (C6) at (\xc+5*\dxc,\yc);

    \filldraw[ABstyle]
        (-4.7,-0.9) -- (-4.8,1.60) --
        (3.5,2.50) -- (3.6,0.00) -- cycle;

    \filldraw[ACstyle]
        (0.4,2.50) -- (0.3,0) --
        (9,-0.9) -- (9.1,1.60) -- cycle;

    \node[Bdot] at (B1) {};
    \node[Bdot] at (B2) {};
    \node[Bdot] at (B3) {};
    \node[Bdot] at (B4) {};
    \node[Bdot] at (B5) {};

    \node[Adot] at (A1) {};
    \node[Adot] at (A2) {};
    \node[Adot] at (A3) {};
    \node[Adot] at (A4) {};

    \node[Cdot] at (C1) {};
    \node[Cdot] at (C2) {};
    \node[Cdot] at (C3) {};
    \node[Cdot] at (C4) {};
    \node[Cdot] at (C5) {};
    \node[Cdot] at (C6) {};

    \node[font=\LARGE,black] at (-1.5,2.6) {$F_{AB}$};
    \node[font=\LARGE,black] at (5.7,2.6) {$G_{AC}$};

    \end{tikzpicture}%
    }

    \caption{Observables $F_{AB}$ and $G_{AC}$ with supports on
    $A\cup B$ and $A\cup C$.}
    \label{fig:observable-overlap}
\end{figure}
\begin{lem}\label{cq-transitiveQ}
    For pairwise disjoint sets $A,B,C\subseteq[n]$ with $|B|=|C|$ and $\kappa=|A|/(|A|+|B|)$, let $\obsf$ and $\obsg$ be observables supported within $A\cup B$ and $A \cup C$ respectively so that $\obsf=\obsf_{AB}\otimes\mI_{\sim(A\cup B)}$ and $\obsg=\obsg_{AC}\otimes\mI_{\sim(A\cup C)}$.
    Suppose a permutation $\pi \in S_n$ satisfies $\pi(i) = i$ for $i \in A$ and $\swap_{\pi} \, \obsf \, \swap_{\pi}^\dagger = \obsg$.
    Then, if $\mathrm{Sym}(\obsf_{AB})$ is transitive on $A\cup B$, we have 
    \begin{align*}
    \left|  \innerwgn{\obsf}{\obsg}\right| & \leq \kappa\normwgn{\obsf}^{2}
         +(1-\kappa)\big| \! \left\langle \obsf , \mI \right\rangle_{\rho^{\otimes n}}\mkern-4mu \big|^{2}.
    \end{align*}
\end{lem}

\begin{proof}
    Expanding $\obsf,\obsg$ in the orthonormal basis $\{\basis{\bs}\}_{\bs\in [d^2]^{n}}$ and using the fact that $\obsf$ and $\obsg$ are supported within $A\cup B$ and $A\cup C$ respectively, we get
\begin{align*}
\obsf &\ =\F{\obsf}_{\mathbf{0}}\basis{\mathbf{0}}+\sum_{\bs\in \cS_{A}}\F{\obsf}_{\bs}\basis{\bs}+\sum_{\bs'\in \cS_{AB}-\cS_{A} }\F{\obsf}_{\bs'}\basis{\bs'}\\
\obsg &\ =\F{\obsg}_{\mathbf{0}}\basis{\mathbf{0}}+\sum_{\bt\in \cS_{A}}\F{\obsg}_{\bt}\basis{\bt}+\sum_{\bt'\in \cS_{AC}-\cS_{A}}\F{\obsg}_{\bt'}\basis{\bt'},
\end{align*}
where we use the shorthand  $\cS_{AB}\coloneqq \cS_{A \cup B}$ and $\cS_{AC} \coloneqq \cS_{A \cup C}$.
The existence of $\pi \in S_n$ such that $\pi(i) = i$ for $i \in A$ and $\swap_{\pi} \, \obsf \, \swap_{\pi}^\dagger = \obsg$ implies that
\begin{align*}
\F{\obsf}_{\mathbf{0}}=\F{\obsg}_{\mathbf{0}}, \quad \text{and} \quad \F{\obsf}_{\bs}=\F{\obsg}_{\bs},\quad \forall \bs\in \cS_{A}.
\end{align*}
Expanding $\innerwgn{\obsf}{\obsg}$ in the orthonormal basis $\{\basis{\bs}\}_{\bs\in [d^2]^{n}}$ and simplifying, we get
\begin{align*}
     &\ \left| \innerwgn{\obsf}{\obsg} \right|\\
     &\ =\left|\innerwgn{\sum_{\bs\in [d^2]^n}\F{\obsf}_{\bs}\basis{\bs}}{\sum_{\bt\in [d^2]^n}\F{\obsg}_{\bt}\basis{\bt}}\right|\\
&\ = 
    \Bigg| \innerwgn{\left(\F{\obsf}_{\mathbf{0}}\basis{\mathbf{0}}+\sum_{\bs\in \cS_{A}}\F{\obsf}_{\bs}\basis{\bs}\right)}{\left(\F{\obsg}_{\mathbf{0}}\basis{\mathbf{0}}+\sum_{\bt\in \cS_{A}}\F{\obsg}_{\bt}\basis{\bt}\right)} \\
&\ \quad + \innerwgn{\sum_{\bs\in \cS_{AB}-\cS_{A}}\F{\obsf}_{\bs}\basis{\bs}}{\sum_{\bt\in \cS_{AC}-\cS_{A}}\F{\obsg}_{\bt}\basis{\bt}}\\
&\ \quad +\innerwgn{\left(\F{\obsf}_{\mathbf{0}}\basis{\mathbf{0}}+\sum_{\bs\in \cS_{A}}\F{\obsf}_{\bs}\basis{\bs}\right)}{\sum_{\bt\in \cS_{AC}-\cS_{A}}\F{\obsg}_{\bt}\basis{\bt}}\\
&\ \quad +\innerwgn{\sum_{\bs\in \cS_{AB}-\cS_{A}}\F{\obsf}_{\bs}\basis{\bs}}{\left(\F{\obsg}_{\mathbf{0}}\basis{\mathbf{0}}+\sum_{\bt\in \cS_{A}}\F{\obsg}_{\bt}\basis{\bt}\right)}\Bigg|\\
&\ =  \left|\innerwgn{\left(\F{\obsf}_{\mathbf{0}}\basis{\mathbf{0}}+\sum_{\bs\in \cS_{A}}\F{\obsf}_{\bs}\basis{\bs}\right)}{\left(\F{\obsf}_{\mathbf{0}}\basis{\mathbf{0}}+\sum_{\bt\in \cS_{A}}\F{\obsf}_{\bt}\basis{\bt}\right)}\right| \\
     &\ = |\F{\obsf}_{\mathbf{0}}|^{2} +\sum_{\bs\in \cS_{A}}|\F{\obsf}_{\bs}|^{2},
\end{align*}
where the cross terms cancel by orthonormality because $A,B,C$ are disjoint and the only nonconstant basis strings that occur in both expansions are those in $\cS_A$.
More specifically, $\innerwgn{\basis{\bs}}{\basis{\bt}}=0$ if $(\bs,\bt) \in (\cS_{AB}-\cS_{A}) \times (\cS_{AC} - \cS_{A})$, $(\bs,\bt) \in \cS_{A} \times (\cS_{AC} - \cS_{A})$, or $(\bs,\bt) \in (\cS_{AB}-\cS_{A}) \times \cS_{A}$. 

For $\bs \in [d^2]^n$ and $i \in [n]$, define the indicator function
\begin{align*}
    Q_{\bs}(i)\coloneqq \begin{cases}
        1 \quad \text{if $s_{i}\neq 0$}\\
        0 \quad \text{if $s_{i}=0$}
    \end{cases}
\end{align*}
and the sums
\begin{align*}
  \eta_{i}(\obsf) &\coloneqq \sum_{\bs\in \cS_{AB}} Q_{\bs}(i) \frac{|\F{\obsf}_{\bs}|^{2}}{|\bs|}, \qquad |\bs|\coloneqq \sum_{i\in [n]} Q_{\bs}(i).
\end{align*}
Then, expanding $\normwgn{\obsf}^{2}$ in the $\{\Omega_{\bs}\}_{\bs\in [d^2]^{n}}$ basis shows that
\begin{align*}
   \normwgn{\obsf}^{2} &\ =|\F{\obsf}_{\mathbf{0}}|^{2} +\sum_{\bs\in \cS_{AB}}|\F{\obsf}_{\bs}|^{2}\\
    &\ =|\F{\obsf}_{\mathbf{0}}|^{2}+\sum_{i\in A\cup B}\sum_{\bs \in \cS_{AB}: s_i \neq 0}\frac{|\F{\obsf}_{\bs}|^{2}}{|\bs|}\\
    &\ = |\F{\obsf}_{\mathbf{0}}|^{2} + \sum_{i\in A\cup B}\eta_{i}(\obsf).
\end{align*}

For any \(\tau\in\mathrm{Sym}(\obsf_{AB})\), we can extend by the identity outside \(A\cup B\) so the extension lies in \(\mathrm{Sym}(\obsf)\) and preserves \(\cS_{AB}\) because \(\obsf=\obsf_{AB}\otimes\mI_{\sim(A\cup B)}\).
Thus, the extended $\tau$ gives
\begin{align*}
    \eta_{\tau(i)}(\obsf) &\ =\sum_{\bs\in \cS_{AB}} Q_{\bs}(\tau(i))\frac{|\F{\obsf}_{\bs}|^{2}}{|\bs|}\\
    &\ \stackrel{(a)}{=} \sum_{\bs\in \cS_{AB}} Q_{\tau \bs}(\tau(i))\frac{|\F{\obsf}_{\tau\bs}|^{2}}{|\tau \bs |}\\
    &\ =\eta_{i}(\obsf),
\end{align*}
where $(a)$ holds because $\tau \cS_{AB} = \cS_{AB}$ holds because $\tau$ must stabilize the support of $\obsf$ and one can change the variable of summation to $\tau \bs$.
Since $\mathrm{Sym}(\obsf_{AB})$ is transitive, one can see that $\eta_i(\obsf)$ does not depend on $i\in A \cup B$ and it follows that
\begin{align*}
    \eta_i (\obsf) &= \bar\eta(\obsf) \coloneqq \frac{1}{|A|+|B|}(\|\obsf\|_{\rho^{\otimes n}}^2-|\widehat \obsf_{\mathbf0}|^2), \; i\in A\cup B.
\end{align*}
 Thus, we get the inequality
\begin{align*}
    \frac{|A|}{|A|+|B|}\left( \normwgn{\obsf}^{2}-|\F{\obsf}_{\mathbf{0}}|^{2}\right) & = |A|\bar\eta(\obsf) \\ &= \sum_{i\in A}\eta_i (\obsf)\\
    & \geq \sum_{i\in A}\sum_{\bs \in \cS_{A}: s_i \neq 0}\frac{|\F{\obsf}_{\bs}|^{2}}{|\bs|}\\
    & =\sum_{\bs\in \cS_{A}}|\F{\obsf}_{\bs}|^{2},
\end{align*}
where the inequality holds because the sum is over $\bs \in \cS_A$ rather than $\bs \in \cS_{AB}$.
Combining the above equations gives the stated result
\begin{align*}
 \left| \innerwgn{\obsf}{\obsg}\right|
   & \leq |\F{\obsf}_{\mathbf{0}}|^{2}+ \frac{|A|}{|A|+|B|}\left( \normwgn{\obsf}^{2}-|\F{\obsf}_{\mathbf{0}}|^{2}\right)\\
& =\kappa \normwgn{\obsf}^{2}
     +(1-\kappa)\big| \! \innerwgn{\obsf}{\mI} \! \big|^{2},
\end{align*}
where the equality holds because \(\F{\obsg}_{\bs}=\F{\obsf}_{\bs}\) for $\bs\in \cS_{A}$.
\end{proof}

While our main result is based on the GNS inner product, most of the ideas (e.g., the above lemma) extend naturally to the Kubo-Martin-Schwinger (KMS) inner product \cite{olkiewicz1999hypercontractivity}. This is discussed in
\ifarxiv
Appendix~\ref{kms}.  
A generalized version of Lemma~\ref{cq-transitiveQ} is established in Appendix~\ref{corr-ineq-lem}.
    \fi

\vspace*{-0mm}

\section{Code Symmetry and Decoder Observables}\label{sec:transitive symm obs}

Consider the RM-code setup from Section~\ref{sec:rm-cq-decoding}, where \(\cC=\RM(r,m)\) and the projections \(\cC'\) and \(\cC''\) both equal \(\RM(r,m-1)\). Here we establish the recursive MMSE comparison between \(\cC\) and these projections. We embed their MMSE observables in the common space \(Y_{\sim0}\), verify the symmetry hypotheses of Lemma~\ref{cq-transitiveQ}, and derive a recursion across blocklengths. Section~\ref{sec:achieving vanishing ber} then combines this recursion with the EXIT bound and the MMSE--Helstrom comparison.


In particular, our goal is to establish two relationships: (i) between $P_{b}(\cC)$ and $ H(W^{\cC}_{\text{extr}})$ and (ii) between $P_{b}(\cC)$ and $P_{b}(\cC')=P_{b}(\cC'')$.
The relationship between $P_{b}(\cC)$ and $H(W^{\cC}_{\text{extr}})$ serves to determine the code rate at which decoding starts to work.
The relationship between $P_{b}(\cC)$ and $P_{b}(\cC')$, $P_{b}(\cC'')$ allows one to relate the Helstrom error probabilities for decoding bit-0 from the extrinsic observation $Y_{\sim 0}$ for $\cC = \RM(r,m)$ with the Helstrom error probabilities for decoding bit-0 from the extrinsic observations of the smaller codes $\cC'$ and $\cC''$.  The remainder of this section establishes the second relationship through an MMSE recursion, while the first relationship is completed in Section~\ref{sec:achieving vanishing
  ber}.


Let $\measo_{\sim 0}$ denote the MSE minimizing observable for the extrinsic channel $W^{\cC}_{\text{extr}}$ used to distinguish $\overline{\rho}^{x}_{Y_{\sim 0}}$ for $x\in \{0,1\}$. Let $\widehat{\measo}_{AB}$ denote the MSE minimizing observable for the extrinsic channel $W^{\cC'}_{\text{extr}}$ used to distinguish $\overline{\rho}^{x}_{Y_{AB}}$ for $x\in \{0,1\}$.
Similarly, let $\widehat{\measo}_{AC}$ denote the observable for $W^{\cC''}_{\text{extr}}$ used to distinguish $\overline{\rho}^{x}_{Y_{AC}}$  for $x\in \{0,1\}$. 
Let $\measo_{AB}$ be the observable $\widehat{\measo}_{AB}$ extended to the full extrinsic space $Y_{\sim 0}$ i.e. $\measo_{AB}=\widehat{\measo}_{AB}\otimes \mI_{Y_{CD}}$.

Using the setup in Proposition~\ref{prop:rm_prop}(c), let $\widetilde\pi \in S_N$ be the permutation that satisfies $\widetilde\pi(i) = i$ for $i\in \{0 \} \cup A \cup D$, $\widetilde\pi(i) = i+2^{m-2}$ for $i\in B$, and $\widetilde\pi(i) = i - 2^{m-2}$ for $i\in C$.
Since \(\widetilde\pi(0)=0\), its restriction to \([N]\setminus\{0\}\), followed by the relabeling $j\mapsto j-1$ (described in Remark~\ref{rem:extrinsic-relabel}), induces a permutation \(\pi\in S_n\).
Hence, in the definition of \(\swap_{BC}\), we use \(\swap_{\pi}\).
This permutation has the effect of swapping the $B$ and $C$ blocks in Figure~\ref{rm-nesting}.
For the unitary swap operator defined by this permutation, we use the shorthand $\swap_{BC}\coloneqq\swap_{\pi}$.
We then define
$\measo_{AC}\coloneqq\swap_{BC}^{\dagger}\measo_{AB}\swap_{BC}$.
Since conjugation by $\swap_{BC}$ maps the extrinsic channel induced by $\cC''$ on $Y_{AC}$ to the extrinsic channel induced by $\cC'$ on $Y_{AB}$, the observable $\measo_{AC}$ is MMSE-achieving for $\cC''$ and has the form
$\measo_{AC}=\widehat{\measo}_{AC}\otimes\mI_{Y_{BD}}$.

These extensions allow comparison of three MMSE quantities using the same output space.
We choose the MMSE-achieving observables to be invariant under codeword symmetries of the corresponding extrinsic channels.
With this, the averaging argument in Lemma~\ref{lem:helstrom-code-symmetry} allows us to evaluate the relevant quadratic losses for the common product state $\meas$. 

Let $\obs_{X_0}=(\ketbra{0}{0}_{X_0}-\ketbra{1}{1}_{X_0})\otimes \mI_{Y_{\sim 0}}$ be the observable for the input $X_0$ to the extrinsic channels. Since we have defined observables $\measo_{AB},\measo_{AC}$ over the full extrinsic space  $Y_{\sim 0}$, the definition of $\obs_{X_0}$ stays the same for codes $\cC$, $\cC'$ and $\cC''$ to analyze extrinsic CQ channels. Similarly we define $\obs_{\sim 0}=\mI_{X_0}\otimes \measo_{\sim 0}$, $\obs_{AB}=\mI_{X_0}\otimes \measo_{AB}$ and $\obs_{AC}=\mI_{X_0}\otimes \measo_{AC}$.

\begin{lem}\label{lem:mmse-observable-transitive-symmetry}
    There is a canonical MMSE-achieving $\widehat{\measo}_{AB}$ invariant under $U^{\bc}$ for $\bc\in\cC'_0$ and the stabilizer $\Gamma_0(\cC')$, antisymmetric under $U^{\otimes|A\cup B|}$, and consequently with transitive symmetry on $A\cup B$.
\end{lem}
\begin{proof}
Define $\rho_{AB}=(\overline\rho^0_{Y_{AB}}+\overline\rho^1_{Y_{AB}})/2$ and $D_{AB}=(\overline\rho^0_{Y_{AB}}-\overline\rho^1_{Y_{AB}})/2$, and let $\widehat{\measo}_{AB}$ be the canonical Sylvester solution on $\supp(\rho_{AB})$, extended by zero on its kernel.
If a unitary $V$ satisfies $V\rho_{AB}V^\dagger=\rho_{AB}$ and $VD_{AB}V^\dagger= \pm D_{AB}$, then the covariance of the Sylvester equation and uniqueness on the support gives $V\widehat{\measo}_{AB}V^\dagger= \pm \widehat{\measo}_{AB}$.
Since $V$ preserves the support, this equality also holds for the zero extension.  
Let $\Gamma_0(\cC')=\{\pi\in\operatorname{Aut}(\cC'): \pi(0)=0\}$.
For $\pi\in\Gamma_0(\cC')$, the induced coordinate permutation preserves each conditional state, while for $\bc\in\cC'_0:=\{\bc\in\cC':c_0=0\}$, $U^{\bc_{AB}}$ also preserves each conditional state.
Hence $\widehat{\measo}_{AB}$ is invariant under both actions.
Double transitivity makes $\Gamma_0(\cC')$ transitive on $A\cup B$, so $\mathrm{Sym}(\widehat{\measo}_{AB})$ is transitive.
The all-ones codeword interchanges the two conditional states, and therefore $U^{\otimes|A\cup B|}\widehat{\measo}_{AB}U^{\otimes|A\cup B|}=-\widehat{\measo}_{AB}$.
Finally, we can extend this observable by the identity on $Y_{CD}$ and define $\measo_{AC}$ by the $B/C$ swap.
\end{proof}

Choose the canonical observable from Lemma~\ref{lem:mmse-observable-transitive-symmetry}.
Its invariance under the zero-coset and stabilizer actions, together with its antisymmetry under the all-ones action, justifies the identities below.

For such an invariant observable $M$, the same averaging argument as in
Lemma~\ref{lem:helstrom-code-symmetry} gives
\begin{align*}
    \Tr\!\left(\overline\rho^0_{Y_{\sim0}}(\mI-M)^2\right)
    =
    \Tr\!\left((\rho^0)^{\otimes n}(\mI-M)^2\right),
\end{align*}
and similarly for the projected extrinsic systems defining $\cC'$ and $\cC''$.
The same averaging identity applies for \(M\) and \(M^2\), since \(M\) is chosen invariant under the action of codewords in the zero coset and therefore \(M^2\) is invariant as well.
Moreover, by the antisymmetric choice from Lemma~\ref{pm-symmetry}, the
MMSE-achieving observable satisfies
\begin{align*}
    \Tr\!\left((\rho^0)^{\otimes n}M\right)
    =
    \Tr\!\left((\rho^0)^{\otimes n}M^2\right)
    =
    1-\mmse(X_0|Y_{\sim0}).
\end{align*}
Indeed, the optimality condition gives
\begin{align*}
    \Tr(\rho_YM^2)
    =
    \Tr(DM)
    =
    1-\mmse(X_0|Y_{\sim0}),
\end{align*}
while antisymmetry gives
\begin{align*}
    \Tr(\rho_YM)=0,
    \qquad
    \Tr(DM^2)=0.
\end{align*}
Since $\rho^0=\rho_Y+D$, the identities follow.
The same arguments extend naturally to the projected extrinsic systems defining $C'$ and $C''$.

Let $\mmse(\cC)$ be the MMSE for the $\RM$ code $\cC$ associated with decoding $X_0$ from the output of the extrinsic CQ channel $W^{\cC}_{\text{extr}}$. 
The MMSE $\mmse(\cC)$ satisfies
\begin{align*}
    \mmse(\cC) &\ = \mmse(X_0|Y_{\sim 0})_{\rho_{X_0 Y_{\sim 0}}}\\
&\ =\Tr\left(\rho_{X_0 Y_{\sim 0}}\left(\obs_{X_0}-\obs_{\sim 0}\right)^{2}\right)\\
    &\ = \Tr\left(\overline{\rho}^{0}_{Y_{\sim 0}}\left(\mI_{Y_{\sim 0}}-\measo_{\sim 0}\right)^{2}\right)\\
    &\ \stackrel{(a)}{=}\Tr\left(\meas\left(\mI_{Y_{\sim 0}}-\measo_{\sim 0}\right)^{2}\right)
\end{align*}
where $(a)$ follows from channel symmetry and the linearity of $\cC$.
On the other hand $\mmse(\cC')$  satisfies 
\begin{align*}
    \mmse(\cC')&\ =\mmse(X_0|Y_{AB})_{\rho_{X_0Y_{AB}}}\\
&\ =\Tr\left(\overline{\rho}^{0}_{Y_{AB}}\left(\mI_{AB}-\widehat{\measo}_{AB}\right)^{2}\right)\\
    &\ \stackrel{(a)}{=}\Tr\left((\overline{\rho}^{0}_{Y_{ABCD}})\left(\mI_{Y_{\sim 0}}-\widehat{\measo}_{AB} \otimes \mI_{CD}\right)^{2}\right)\\
    &\ \stackrel{(b)}{=}\Tr\left(\meas \left(\mI_{Y_{\sim 0}}-\measo_{AB} \right)^{2}\right)
\end{align*}
where $(a)$ holds because $\mse(\rho_{X_0 Y_{ABCD}},\widehat{\measo}_{AB} \otimes \mI_{CD})=\mse(\rho_{X_0 Y_{AB}},\widehat{\measo}_{AB})$ and $(b)$ follows from $\measo_{AB}=\widehat{\measo}_{AB}\otimes \mI_{CD}$ and the reduction to the all-zero codeword due to symmetry.
Similarly, $\mmse(\cC'')$ satisfies
\begin{align*}
  \mmse(\cC'') &\ =\mmse(X_0|Y_{AC})_{\rho_{X_0Y_{AC}}}\\
&\ =\Tr\left(\meas \left(\mI_{Y_{\sim 0}}-\widehat{\measo}_{AC} \otimes \mI_{BD}\right)^{2}\right)\\
&\ = \Tr\left(\meas \left(\mI_{Y_{\sim 0}}-\measo_{AC} \right)^{2}\right).
\end{align*}

Since both $\cC'$ and $\cC''$ are $\RM( r,m-1)$ codes, it follows that $\mmse(\cC')=\mmse(\cC'')$.
On the other hand, Lemma~\ref{mmse-p-error} implies that $\mmse(\cC)$ and $\mmse(\cC')$ satisfy 
\begin{alignat*}{3}
    2 P_b(\cC) & \leq \; & \mmse(\cC)  & \leq \;\; &  4P_b(\cC)(1-P_b(\cC)) \;\;\; \\
    2 P_b(\cC') &\leq \; &\mmse(\cC') & \leq \;\; &  4P_{b}(\cC')(1-P_b(\cC')).
\end{alignat*}
Inspired by the analysis of $\RM$ codes over classical BMS channels via the two-look approach in~\cite{Pfister-arxiv25a}, we consider an extension to BSCQ channels.
In particular, we derive a recursive relationship for the extrinsic MMSE of $\RM$ codes on BSCQ channels by analyzing the observables $\measo_{AB}$, $\measo_{AC}$ and $\measo_{\sim 0}$. For this analysis the primary quantity of interest is $ \innerw{\measo_{AB}}{\measo_{AC}} $. Although the analysis of $ \innerw{\measo_{AB}}{\measo_{AC}} $ is somewhat involved, we will now show that it puts us on the right path to obtain a recursive bound for the MMSE in this setting.

\begin{rem}
The observable $\widehat{\measo}_{AB}$ may be chosen to be the MMSE-achieving observable obtained in Lemma~\ref{lem:mmse-observable-transitive-symmetry}.
Consequently, the extended observable $\measo_{AB}$ satisfies that $\mathrm{Sym}(\measo_{AB})$ is transitive on $A\cup B$, while the corresponding observable $\measo_{AC}$ remains MMSE-achieving for $\cC''$.
\end{rem}

Now, we verify the hypotheses of Lemma~\ref{cq-transitiveQ}.
The observable $\measo_{AB}$ is supported within $A\cup B$, while $\measo_{AC}$ is supported within $A\cup C$.
The swap of $B$ and $C$ fixes every coordinate in $A$ and maps one projected extrinsic channel to the other.  By Lemma~\ref{lem:mmse-observable-transitive-symmetry}, $\measo_{AB}$ may be chosen so that $\mathrm{Sym}(\measo_{AB})$ is transitive on $A\cup B$.  Therefore, we can apply Lemma~\ref{cq-transitiveQ} with $\kappa=|A|/(|A|+|B|)$.
For the RM partition in Proposition~\ref{prop:rm_prop}, we have $\kappa\leq 1/2$.
Since $\normw{\measo_{AB}}^2=1-\mmse(\cC')$ is at least $|\innerw{\measo_{AB}}{\mI}|^2=(1-\mmse(\cC'))^2$ and $\kappa\leq1/2$, the resulting correlation bound is at most
\begin{align*}
    \left| \innerw{\measo_{AB}}{\measo_{AC}} \mkern-1mu \right|   &\ \leq \frac{1}{2} \normw{\measo_{AB}}^{2}\\
        &\ \quad + \frac{1}{2} \innerw{\measo_{AB}}{\mI}^2,
\end{align*}
where the last term is real because $\measo_{AB}$ is Hermitian.

The previous bound controls the only term that depends on the interaction between the two projected observables.
Since $\measo_{\sim0}$ is optimal for the full extrinsic channel, testing any observable on $Y_{\sim0}$ gives an upper bound on $\mmse(\cC)$.
For the test observable $\alpha(\measo_{AB}+\measo_{AC})$, expand the resulting quadratic loss under $\meas$, and then optimize the scalar $\alpha$.
This gives a recursive upper bound on $\mmse(\cC)$ in terms of $\mmse(\cC')$.

\begin{lem}\label{recursive-cq}
For the $\RM$ code $\cC=\RM(r,m)$ and disjoint sets
$A,B,C\subseteq[N]\setminus\{0\}$ such that the two code projections
\[
    \cC'=\cC|_{\{0\}\cup A\cup B},
    \qquad
    \cC''=\cC|_{\{0\}\cup A\cup C}
\]
are equal to $\RM(r,m-1)$, if we have $\mmse(\cC ')\in [0,1)$, then the extrinsic MMSEs satisfy
\[
    \frac{\mmse(\cC)}{1-\mmse(\cC)}
    \leq
    \frac{3}{4}
    \frac{\mmse(\cC')}{1-\mmse(\cC')}.
\]
\end{lem}

\begin{proof}
From the definition of $\mmse(\cC)$ we get 
    \begin{align*}
     \mmse(\cC) &\ =  \mse(\rho_{X_0 Y_{\sim 0}},\measo_{\sim 0})\\
      &\ = \Tr\left(\meas\left(\mI_{Y_{\sim 0}}-\measo_{\sim 0}\right)^{2}\right)\\
       &\ \stackrel{(a)}{\leq} \Tr \left(\meas \left(\mI_{Y_{\sim 0}}-\alpha(\measo_{AB}+\measo_{AC}) \right)^{2} \right)\\
      &\ = 1-2\alpha \Tr\left(\meas (\measo_{AB}+\measo_{AC})\right) \\ &\ \quad +\alpha^2 \Tr\left(\meas (\measo_{AB}+\measo_{AC})^2\right)\\  
      &\ = 1-4\alpha(1-\mmse(\cC')) +2 \alpha^2(1-\mmse(\cC')) \\ &\ \quad +2\alpha^2 \Re \big( \Tr\left(\measo_{AB}\meas \measo_{AC}\right) \big),
    \end{align*}
    where $(a)$ is true because $\measo_{\sim 0}$ is the optimal MMSE observable.
    Using Lemma~\ref{cq-transitiveQ}, we can bound the cross term with
\[
    \left|\innerw{\measo_{AB}}{\measo_{AC}}\mkern-1mu \right|
    \leq
    \kappa(1-\mmse(\cC'))+(1-\kappa)(1-\mmse(\cC'))^2,
\]
where $\kappa$ is specified later.
This gives
\begin{align*}
    \mmse(\cC)
    &\leq
    1+2\alpha(\alpha-2)(1-\mmse(\cC'))\\
    &\quad
    +2\alpha^2\left[\kappa(1-\mmse(\cC'))+(1-\kappa)(1-\mmse(\cC'))^2\right]\\
    &=
    1-2(1-\mmse(\cC'))
    \left[2\alpha-\alpha^2\big(2-(1-\kappa)\mmse(\cC')\big)\right].
\end{align*}
The right-hand side is minimized by choosing
\[
    \alpha^*=\frac{1}{2-(1-\kappa)\mmse(\cC')},
\]
and this results in the bound
\[
    \mmse(\cC)
    \leq
    \frac{(1+\kappa)\mmse(\cC')}{2-(1-\kappa)\mmse(\cC')}.
\]
Applying the increasing map \(\phi(x)=x/(1-x)\) on \([0,1)\) yields
\[
    \frac{\mmse(\cC)}{1-\mmse(\cC)}
    \leq
    \frac{1+\kappa}{2}
    \frac{\mmse(\cC')}{1-\mmse(\cC')}.
\]
For the given RM setup, we have \(\kappa\leq 1/2\) and this gives the stated result.
\end{proof}

\section{Achieving Vanishing Bit-Error Probability}\label{sec:achieving vanishing ber}

In this section, we establish the primary result on vanishing bit error probability by combining the results from previous sections.
Lemma~\ref{rate-relation-cq} gives an initial upper bound on the extrinsic entropy when the code rate is below the Holevo capacity.
Shortly, we will see that Lemma~\ref{mmse-entropy} converts this entropy bound into an initial upper bound on the MMSE.
Lemma~\ref{recursive-cq} then propagates the bound through the nested RM sequence, and Lemma~\ref{mmse-p-error} finally converts the resulting MMSE bound into a bit-error probability bound.

From our definition of extrinsic channel $W^{\cC}_{\text{extr}}$ of code $\cC$ associated with CQ state $\rho_{X_0Y_{\sim 0}}$ we have
\begin{align*}
    H(X_0|Y_{\sim 0})_{\rho_{X_0\!Y_{\sim 0}}}= H(W^{\cC}_{\text{extr}}).
\end{align*}
For $W^{\cC}_{\text{extr}}$, we can revisit the relationship between the Helstrom error probability and the conditional min-entropy $H_{\min}(W^{\cC}_{\text{extr}}) = H_{\min}(X_0 |Y_{\sim 0})_{\rho_{X_0\!Y_{\sim 0}}}$ to see that 
\[ 1 - P_e \left(\overline{\rho}^{0}_{Y_{\sim 0}}, \overline{\rho}^{1}_{Y_{\sim 0}},\frac{1}{2} \right) = 2^{-H_{\min}(X_0|Y_{\sim 0})_{\rho_{X_0\!Y_{\sim 0}}} }. \]
The conditional min-entropy also satisfies the inequality
\[ H(X_0|Y_{\sim 0})_{\rho_{X_0\!Y_{\sim 0}}} \geq H_{\min} (X_0|Y_{\sim 0})_{\rho_{X_0\!Y_{\sim 0}}} \]
and this implies that
\[  P_e \left(\overline{\rho}^{0}_{Y_{\sim 0}}, \overline{\rho}^{1}_{Y_{\sim 0}},\frac{1}{2} \right) \leq 1- 2^{-H (X_0|Y_{\sim 0})_{\rho_{X_0Y_{\sim 0}}}}. \]

Thus, an upper bound on $H(X_0|Y_{\sim0})_{\rho_{X_0Y_{\sim0}}}$ gives an upper bound on the Helstrom error probability of the extrinsic channel.
Applying Lemma~\ref{mmse-p-error} then gives the corresponding upper bound on the MMSE.
The next lemma states this conversion in the form needed to initialize the RM recursion.

\begin{lem}\label{mmse-entropy}
 The MMSE of $X_0$ from the observations of the quantum systems $Y_{\sim 0}$ satisfies
 \begin{align*}
&\ \mmse{\left(X_0\mid Y_{\sim 0}\right)}_{\rho_{X_0Y_{\sim 0}}} \\
 &\ \leq 4 \cdot 2^{-H (X_0|Y_{\sim 0})_{\rho_{X_0Y_{\sim 0}}}} \big(1-2^{-H (X_0|Y_{\sim 0})_{\rho_{X_0Y_{\sim 0}}}}\big).
\end{align*}
Thus, if $H (X_0|Y_{\sim 0})_{\rho_{X_0Y_{\sim 0}}}\leq 1-\frac{\delta}{\ln 2}$ for $\delta\in [0,\ln 2]$, then the MMSE satisfies $\mmse{\left(X_0\mid Y_{\sim 0}\right)}_{\rho_{X_0Y_{\sim 0}}} \leq 1-\delta^2$ and
\[ \mmse{\left(X_0\mid Y_{\sim 0}\right)}_{\rho_{X_0Y_{\sim 0}}}  \!\leq 1-(\ln 2)^2(1-H (X_0|Y_{\sim 0})_{\rho_{X_0Y_{\sim 0}}})^2. \]
\end{lem}

\begin{proof}
We know that
\[ P_e \left(\overline{\rho}^{0}_{Y_{\sim 0}},\overline{\rho}^{1}_{Y_{\sim 0}},\frac{1}{2} \right) = \frac{1-T}{2}, \]
where
\[ T = \frac{1}{2} \|  \overline{\rho}^{0}_{Y_{\sim 0}}- \overline{\rho}^{1}_{Y_{\sim 0}} \|_1 \geq 1-2 \left(1-2^{-H (X_0|Y_{\sim 0})_{\rho_{X_0Y_{\sim 0}}} } \right). \]
Thus, we find that
\begin{align*}
&\ \mmse{\left(X_0\mid Y_{\sim 0}\right)}_{\rho_{X_0Y_{\sim 0}}}\\
&\leq  4P_e \left(\overline{\rho}^{0}_{Y_{\sim 0}},\overline{\rho}^{1}_{Y_{\sim 0}},\frac{1}{2} \right) \left(1-P_e \left( \overline{\rho}^{0}_{Y_{\sim 0}},\overline{\rho}^{1}_{Y_{\sim 0}},\frac{1}{2} \right)\right) \\
&\leq 1-T^2 \\
&\leq 4 \cdot 2^{-H (X_0|Y_{\sim 0})_{\rho_{X_0Y_{\sim 0}}}} \big(1-2^{-H (X_0|Y_{\sim 0})_{\rho_{X_0Y_{\sim 0}}}}\big).
\end{align*}
For the last part, the upper bound on the MMSE monotonically increases to 1 as the conditional entropy increases to 1.
By choosing $H (X_0|Y_{\sim 0})_{\rho_{X_0Y_{\sim 0}}} = 1-\frac{\Delta}{\ln 2}$ for $\Delta \in [0,\ln 2]$, we see that the RHS expression equals
\[ 2e^\Delta - e^{2\Delta} = \sum_{k=0}^\infty \frac{2\Delta^k - 2^k \Delta^k}{k!} \leq 1-\Delta^2 \]
using the $k=0,1,2$ terms because the rest are negative.
This implies the final claim that
\[ \mkern-1.5mu \mmse(X_0|Y_{\sim 0})_{\rho_{X_0Y_{\sim 0}}} \!\leq 1 \mkern-0.5mu - \mkern-0.5mu (\ln 2)^2(1 \mkern-1.5mu - \mkern-1.5mu H (X_0|Y_{\sim 0})_{\rho_{X_0Y_{\sim 0}}})^2.\! \qedhere\]

\end{proof}

Lemma~\ref{mmse-entropy} gives the initial condition for the base code $\cC_0$ once $R(\cC_0)<C$.
The recursive inequality from Lemma~\ref{recursive-cq} is then applied repeatedly along the sequence $\cC_k=\RM(r,m+k)$.
It is convenient to iterate the recursion after applying the increasing map $x\mapsto x/(1-x)$, which turns the recursion into a geometric bound.

\begin{lem}\label{error-rate}
Consider the sequence of $\RM$ codes such that $\cC_0=\RM(r_m,m)$ and
$\cC_k=\RM(r_m,m+k)$.  If $\mmse(\cC_0)\leq 1-\delta$ for $\delta>0$, then, for all $k\geq0$, we have
\[
    \frac{\mmse(\cC_k)}{1-\mmse(\cC_k)}
    \leq
    \left(\frac{3}{4}\right)^k\frac{1-\delta}{\delta}.
\]
Equivalently, this gives
\[
    \mmse(\cC_k)
    \leq
    \frac{(1-\delta)(3/4)^k}{\delta+(1-\delta)(3/4)^k}
    \leq
    \frac{1-\delta}{\delta}\left(\frac{3}{4}\right)^k.
\]
\end{lem}

\begin{proof}
By Lemma~\ref{recursive-cq}, for every \(j\geq0\),
\[
    \frac{\mmse(\cC_{j+1})}{1-\mmse(\cC_{j+1})}
    \leq
    \frac{3}{4}
    \frac{\mmse(\cC_j)}{1-\mmse(\cC_j)}.
\]
Iterating from \(\mmse(\cC_0)\leq1-\delta\) shows that
\[
    \frac{\mmse(\cC_k)}{1-\mmse(\cC_k)}
    \leq
    \left(\frac{3}{4}\right)^k\frac{1-\delta}{\delta}.
\]
Solving this inequality for \(\mmse(\cC_k)\) gives
\[
    \mmse(\cC_k)
    \leq
    \frac{(1-\delta)(3/4)^k}{\delta+(1-\delta)(3/4)^k}
    \leq
    \frac{1-\delta}{\delta}\left(\frac{3}{4}\right)^k. \qedhere
\]
\end{proof}

Now, we can state the main quantitative theorem.
The theorem combines the RM rate estimate for the sequence $\cC_k=\RM(r,m+k)$, the initial MMSE bound obtained from the rate gap $C-R(\cC_0)$, and the recursive decay from Lemma~\ref{error-rate}.
The last step uses Lemma~\ref{mmse-p-error} to pass from $\mmse(\cC_k)$ to $P_b(\cC_k)$.

\begin{theorem}\label{error-rate-exponent}
Consider a BSCQ channel $W$ with Holevo capacity $C$.
For any $r,m\in\mathbb{N}$ with $r<m$, consider the $\RM$ code sequence
$\cC_k=\RM(r,m+k)$.
Then \(R(\cC_k)\geq R(\cC_0)-k/(2\sqrt m)\) and, if \(R(\cC_0)<C\), we have
\[
    P_b(\cC_k)
    \leq
    \frac{1}{2}
    \frac{(1-\delta)(3/4)^k}{\delta+(1-\delta)(3/4)^k}
    \leq
    \frac{1-\delta}{2\delta}\left(\frac{3}{4}\right)^k,
\]
where \(\delta=(\ln2)^2(C-R(\cC_0))^2>0\).
\end{theorem}
\begin{proof}
The code rate satisfies
\[
    R(\cC_k)\geq R(\cC_0)-\frac{k}{2\sqrt m},
\]
for all \(k\in\mathbb{N}_0\)~\cite{Pfister-arxiv25a}.
Next, Lemmas~\ref{rate-relation-cq} and~\ref{mmse-entropy} imply
\begin{align*}
    \mmse(\cC_0)
    &=
    \mmse(X_0|Y_{\sim0})_{\rho_{X_0Y_{\sim0}}}\\
    &\leq
    1-(\ln2)^2
    \big(1-H(X_0|Y_{\sim0})_{\rho_{\bX\!\bY}}\big)^2\\
    &\leq
    1-(\ln2)^2(C-R(\cC_0))^2,
\end{align*}
where $C$ is the BSCQ channel capacity.
Thus, we can choose \(\delta=(\ln2)^2(C-R(\cC_0))^2\) and apply Lemma~\ref{error-rate} to see that
\[
    \mmse(\cC_k)
    \leq
    \frac{(1-\delta)(3/4)^k}{\delta+(1-\delta)(3/4)^k}.
\]
Finally, Lemma~\ref{mmse-p-error} gives \(2P_b(\cC_k)\leq\mmse(\cC_k)\), and hence
\[
    P_b(\cC_k)
    \leq
    \frac{1}{2}
    \frac{(1-\delta)(3/4)^k}{\delta+(1-\delta)(3/4)^k}
    \leq
    \frac{1-\delta}{2\delta}\left(\frac{3}{4}\right)^k. \qedhere
\]
\end{proof}

The theorem is stated for a fixed base code and a sequence obtained by increasing the blocklength parameter.  To obtain the informal asymptotic statement, one chooses the base code rate below $C$ by a fixed amount and then lets $k$ grow proportionally to $\sqrt{m}$.  The RM rate estimate controls the loss in rate, while the factor $(3/4)^k$ gives exponential decay in $\sqrt{m}$.
Formalizing this argument gives Theorem~\ref{informal-main-result}.

\begin{proof}[Proof of Theorem~\ref{informal-main-result}]
To recover Theorem~\ref{informal-main-result} from Theorem~\ref{error-rate-exponent}, we use the rate analysis ideas from~\cite[p.~924]{Reeves-it23}.
First, for large enough $m$, fix a target rate gap \(\eta>0\) and choose \(k=\lfloor (\eta/2)\sqrt m\rfloor\) and $m'=m-k$.
As $m$ is large enough, we assume $k/(2\sqrt{m'}) \leq \eta/3$.
Since increasing $r$ changes the rate by $2^{-m'}\binom{m'}{r}$, whose maximum is $O(1/\sqrt{m'})$, there is an $r\in \mathbb{N}$ such that \(\cC_0=\RM(r,m')\) satisfies \(\eta/4<C-R(\cC_0)\leq \eta/2\).
Apply Theorem~\ref{error-rate-exponent} to get \(C-R(\cC_k)\leq \eta\), \(\delta \geq(\ln2)^2(\eta/4)^2\), and
\[
    P_b(\cC_k)
    \leq
    \frac{1-\delta}{2\delta}
    \exp\!\left(-\frac{1}{2}\ln(4/3) \eta\sqrt m + O (1)\right).
\]
Thus, for any \(c<\frac{1}{2}\ln(4/3)\), there is an \(m_0\) such that, for all \(m>m_0\), \(P_b(\cC_k)\leq e^{-c\eta\sqrt m}\).
\end{proof}

The bound above is for one fixed coordinate.
Since RM codes are transitive, the same bit-error bound holds for every coordinate.
Therefore, for any chosen coordinate set, we can apply the corresponding single-coordinate projective Helstrom measurements sequentially to the received state.
Gao's quantum union bound controls the cumulative error despite the disturbance caused by earlier measurements~\cite{gao2015quantum}.

\begin{lem}
Let \(S\subseteq[N]\).  For each \(i\in S\), let \(\Pi_{i,+},\Pi_{i,-},\Pi_{i,\circ}\) be the spectral projectors of the \(i\)-th extrinsic Helstrom difference, choose any projector \(Q_i\leq\Pi_{i,\circ}\), and form the optimal binary projective completion \(\Pi_{i,0}=\Pi_{i,+}+Q_i\), \(\Pi_{i,1}=\Pi_{i,-}+(\Pi_{i,\circ}-Q_i)\), extended by the identity on \(Y_i\).  For a transmitted codeword \(\bx\), Gao's union bound gives
\[ \Pr\{\text{Error on any bit in $S$}\mid\bx\}\leq4\sum_{i\in S}\Tr((\mI-\Pi_{i,x_i})\rho_{\bY}^{\bx}). \]
Averaging over a uniformly transmitted codeword gives
\[ \Pr\{\text{Error on any bit in $S$}\} \leq4\sum_{i\in S}P_{b,i}(\cC)=4|S|P_b(\cC), \]
where the last step uses transitivity.
Since \(N=2^m\) and \(P_b(\cC)\leq e^{-c\eta\sqrt m}\), this tends to zero when \(|S|=2^{o(\sqrt{\log N})}\).
\end{lem}

\section{Conclusion}

While $\RM$ codes have been studied extensively on BMS channels, comparatively little is known about their behavior on classical--quantum channels. In this paper, we develop a framework for analyzing $\RM$ codes over CQ channels through their bitwise optimal measurements. Although our primary goal was to establish vanishing bit-error probability for $\RM$ codes, the proof required a new correlation bound for quantum observables with suitable support and symmetry properties. We also developed an MMSE formulation for estimating a binary classical variable from a quantum system. Using the correlation bound, we showed that, when the code rate has a fixed gap below the Holevo capacity as the length increases, the MMSE and bit-error probability of each coordinate decay as $e^{-c\sqrt{\log N}}$. Together with the quantum union bound~\cite{gao2015quantum}, this implies that any set of $2^{o(\sqrt{\log N})}$ coordinates can be decoded sequentially with vanishing error probability.

The principal remaining question is whether this result can be strengthened to vanishing block-error probability at all rates below the Holevo capacity. Such an extension would be significant beyond classical--quantum coding. 
Binary pure-state channels provide a particularly compelling test case for such an extension. Through classical--quantum channel duality, a capacity-achieving block-decoding theorem in this setting would translate into a strong-secrecy result for dual Reed--Muller constructions on the corresponding binary symmetric wiretap channel~\cite{renes2017duality,Rengaswamy-isit21}. Existing work comes close to this goal but does not yet establish secrecy capacity in full generality~\cite{Pathegama_2023}. Thus, developing techniques that bridge bitwise optimality and collective block decoding is a natural next step, with consequences for both quantum decoding and classical wiretap coding.

\section*{Acknowledgements}

H.\ Pfister would like to acknowledge useful discussions with Galen Reeves during this research.
The authors made use of generative AI tools, including ChatGPT, to assist in improving the clarity and presentation of the manuscript. All mathematical arguments, results, and interpretations were independently checked and validated by the authors, who take full responsibility for the content of the manuscript.

\begin{appendices}

\renewcommand\thesubsection{\thesection.\Roman{subsection}}
\renewcommand\thesubsectiondis{\thesection.\Roman{subsection}}

\section{Properties of MMSE}\label{property mmse}

\begin{defn} For states $\rho,\sigma\in\dop^d$ with $\mathrm{supp}(\rho)\subseteq\mathrm{supp}(\sigma)$, define the symmetric logarithmic derivative (SLD) quantum $\chi^2$ divergence to be
\[
    \chi^2_{\mathrm{SLD}}(\rho\|\sigma)
    \coloneqq
    \Tr\!\left((\rho-\sigma)\Omega_{\sigma}^{-1}(\rho-\sigma)\right),
\]
where
$\Omega_{\sigma}(A)\coloneqq\frac{1}{2}(\sigma A+A\sigma)$ is the SLD Fisher information operator and $\Omega_{\sigma}^{-1}$ denotes the inverse on the support of $\sigma$.
If the support condition fails, set $\chi^2_{\mathrm{SLD}}(\rho\|\sigma)=\infty$.
When $\rho$ and $\sigma$ commute, this reduces to the classical $\chi^2$ divergence
\[
    \chi^2(P\|Q)=\sum_{y:Q(y)>0}\frac{(P(y)-Q(y))^2}{Q(y)}
\]
for their common eigenvalue distributions $P$ and $Q$.
\end{defn}

\begin{lem}\label{mmse-sld-chi-square}
For the uniform binary CQ state
\[
    \rho_{XY}
    =
    \frac{1}{2}\ketbra{0}{0}_X\otimes\rho^0_Y
    +
    \frac{1}{2}\ketbra{1}{1}_X\otimes\rho^1_Y,
\]
let $\bar{\rho}=(\rho^0+\rho^1)/2$.
Then, we have
\[
    1-\mmse(X|Y)_{\rho_{XY}}
    =
    \chi^2_{\mathrm{SLD}}(\rho^0\|\bar{\rho}).
\]
When $\rho^0$ and $\rho^1$ commute, the same identity holds for the classical $\chi^2$ divergence between their eigenvalue distributions $P_0$ and $P_1$ with $\bar{P}=(P_0+P_1)/2$:
\begin{align*}
    1-\mmse(X|Y)
    &=
    \chi^2(P_0\|\bar{P}).
\end{align*}
\end{lem}

\begin{proof}
For the uniform input distribution, $\rho_Y=\bar{\rho}$ and
$D=(\rho^0-\rho^1)/2=\rho^0-\bar{\rho}$.
By Lemma~\ref{mmse-observable}, the optimal observable satisfies
$\Omega_{\bar{\rho}}(\measo_Y^*)=D$, and
\begin{align*}
    \mmse(X|Y)_{\rho_{XY}}
    &=
    1-\Tr(D\measo_Y^*)
    =
    1-\Tr\!\left(D\Omega_{\bar{\rho}}^{-1}(D)\right)
    \\
    &=
    1-\chi^2_{\mathrm{SLD}}(\rho^0\|\bar{\rho}).
\end{align*}
Since $\rho^1-\bar{\rho}=-D$, the same formula holds with $\rho^1$ in place of $\rho^0$.
When the states commute, diagonalizing them simultaneously turns \(\Omega_{\bar\rho}^{-1}\) into entrywise division by \(\bar P\), which results in \(\chi^2(P_0\Vert\bar P)\).
\end{proof}

\section{KMS Inner Product and Canonical Basis}\label{kms}
    This appendix records the KMS analogue of the GNS construction for comparison but it is not needed for the proof of the main coding theorem.
    For $\rho \in \pd^d$ and $\obsf,\obsg \in \hop^d$, the KMS inner product on $\hop^d$ is denoted by
\[ \inner{\obsf}{\obsg}_\rho^{\mathrm{KMS}} \triangleq \Tr(\obsg^{\dagger} \rho^{\frac{1}{2}} \obsf\rho^{\frac{1}{2}}). \]

 By the cyclic property of the trace, the KMS inner product is always real for $\obsf,\obsg \in \hop^d$.
Thus, the set $\hop^d$ with this inner product is a Hilbert space over $\mathbb{R}$.
This space has dimension $d^2$ because each element of $\hop^d$ is defined by exactly $d^2$ real parameters.

For the KMS inner product, one can find an orthonormal basis $\{ \basis{0}', \basis{1}', \ldots, \basis{d^2 -1}' \}$ for the vector space $\hop^d$ over $\mathbb{R}$, where $\basis{i}' \in \hop^d$ for $i \in [d^2]$ and $\basis{0}' = \mI$.
For example, one can apply Gram-Schmidt to any basis whose first element is $\mI$.
Choosing $\basis{0}'=\mI$ guarantees that
\[\inner{ \obsf }{ \basis{0}' }_\rho^{\mathrm{KMS}} = \inner{\obsf}{\mI}_\rho^{\mathrm{KMS}} = \Tr(\rho \obsf).\]

For $\obsf,\obsg \in \hop^{d^n}$ satisfying $\obsf=\otimes_{i=0}^{n-1} \obsf_i$ and $\obsg=\otimes_{i=0}^{n-1} \obsg_i$, one can define
\[ \inner{\obsf}{\obsg}_{\rho,n}^{\mathrm{KMS}} \triangleq  \prod_{i=0}^{n-1} \inner{\obsf_i}{\obsg_i}_{\rho}^{\mathrm{KMS}} = \prod_{i=0}^{n-1} \Tr( \obsg_i^{\dagger} \rho^{\frac{1}{2}}\obsf_i\rho^{\frac{1}{2}}). \]
Similar to the GNS inner product, the extension to the full space satisfies 
\[ \innerwgn{\obsf}{\obsg}^{\mathrm{KMS}} = \Tr(\obsg^{\dagger} (\measgn)^{\frac{1}{2}} \obsf (\measgn)^{\frac{1}{2}}) \]
and the single-system orthonormal basis naturally extends to an orthonormal basis for $\hop^{d^n}$ using the inner product on the full space.
Specifically, for $\bs \in [d^2]^n$, we define the basis elements
\[ \basis{\bs}' \triangleq \basis{s_0}' \otimes \cdots \otimes \basis{s_{n-1}}' \]
and observe that
\begin{align*}
    \innerwgn{\basis{\bs}'}{\basis{\bt}'}^{\mathrm{KMS}} &\ = \prod_{i=0}^{n-1} \Tr( \basis{t_i}'^{\dagger}(\rho)^{\frac{1}{2}}\basis{s_i}'(\rho)^{\frac{1}{2}} )\\
    &\ = \prod_{i=0}^{n-1} \delta_{s_i,t_i}  = \delta_{\bs,\bt}
\end{align*}

For an observable $\obsf\in \hop^{d^n}$, let $\widehat{\obsf}_{\bs}^{\mathrm{KMS}} = \innerwgn{\obsf}{\basis{\bs}'}^{\mathrm{KMS}}$ so that we get the KMS decomposition for $\obsf$  as
\[ \obsf = \sum_{\bs \in [d^2]^n} \widehat{\obsf}_{\bs}^{\mathrm{KMS}} \basis{\bs}'. \]
On the other hand, it also satisfies Parseval's theorem as
\begin{align*}
    \left(\normwgn{\obsf}^{\mathrm{KMS}}\right)^{2} &\ = \Tr(\obsf^{\dagger} (\measgn)^{\frac{1}{2}} \obsf (\measgn)^{\frac{1}{2}})\\
    &\ = \sum_{\bs \in [d^2]^n} |\widehat{\obsf}_{\bs}^{\mathrm{KMS}}|^2 .
\end{align*}
Thus, we find that the variance relation 
\begin{align*}
    \text{Var}_{\mathrm{KMS}}(\obsf)_{\measgn}= &\ \left(\normwgn{\obsf}^{\mathrm{KMS}}\right)^{2}- \left(\innerwgn{\obsf}{\mI}^{\mathrm{KMS}}\right)^{2}\\
    &\ =  \sum_{\bs \in [d^2]^n : \bs \neq \bm{0}} |\widehat{\obsf}_{\bs}^{\mathrm{KMS}}|^2 .
\end{align*}
Using these properties, we can also obtain the transitive symmetry lemma for observables $\obsf$ and $\obsg$ under the KMS inner product.
\begin{lem}\label{cq-transitivekms}
    For pairwise disjoint sets $A,B,C\subseteq[n]$ with $|B|=|C|$ and $\kappa=|A|/(|A|+|B|)$, let $\obsf$ and $\obsg$ be observables supported within $A\cup B$ and $A \cup C$ respectively so that $\obsf=\obsf_{AB}\otimes\mI_{\sim(A\cup B)}$ and $\obsg=\obsg_{AC}\otimes\mI_{\sim(A\cup C)}$.
    Suppose a permutation $\pi \in S_n$ satisfies $\pi(i) = i$ for $i \in A$ and $\swap_{\pi} \, \obsf \, \swap_{\pi}^\dagger = \obsg$.
    Then, if $\mathrm{Sym}(\obsf_{AB})$ is transitive on $A\cup B$, we have
    \begin{align*}
      \left|\innerwgn{\obsf}{\obsg}^{\mathrm{KMS}}\right| & \leq \kappa\left(\normwgn{\obsf}^{\mathrm{KMS}}\right)^{2}
         +(1-\kappa)\left( \innerwgn{\obsf}{\mI}^{\mathrm{KMS}}\right)^2.
\end{align*}
\end{lem}

\begin{proof}
The proof of Lemma~\ref{cq-transitiveQ} applies to this case simply by changing all GNS inner products and norms to KMS inner products and norms.
\end{proof}

\section{Correlation Inequality Lemma}\label{corr-ineq-lem}

For completeness, we now give a two-observable GNS correlation bound that does not require one observable to be a permuted copy of the other.  It strengthens the structural viewpoint of Lemma~\ref{cq-transitiveQ} but is not used in the main recursion.
\begin{lem} \label{lem:corr-ineq}
     Consider disjoint sets $A,B,C\subseteq[n]$ with  $\kappa_1=\frac{|A|}{|A|+|B|}$ and $\kappa_2=\frac{|A|}{|A|+|C|}$. Let $\obsf$ and $\obsg$ be observables on the full $ABC$ system supported $A\cup B$ and $A \cup C$, respectively. Then, if $\mathrm{Sym}(\obsf_{AB})$ is transitive on $A\cup B$ and $\mathrm{Sym}(\obsg_{AC})$ is transitive on $A\cup C$, then we have \vspace*{-1.5mm}
    \begin{align*}
      \left|\innerwgn{\obsf}{\obsg}\right| & \leq \sqrt{\kappa_1\normwgn{\obsf}^{2}
         +(1-\kappa_1)\big| \! \left\langle \obsf , \mI \right\rangle_{\rho^{\otimes n}}\! \big|^{2}}\\
         & \quad \times \sqrt{\kappa_2\normwgn{\obsg}^{2}
         +(1-\kappa_2)\big| \mkern-4mu \left\langle \obsg , \mI \right\rangle_{\rho^{\otimes n}}\mkern-4mu \big|^{2}}.
    \end{align*}
\end{lem}

\begin{proof}
Using the expansion of $\obsf,\obsg$ in the orthonormal basis $\{\basis{\bs}\}_{\bs\in [d^2]^{n}}$ as shown in the proof of Lemma~\ref{cq-transitiveQ}, we get
\begin{align*}
   &\ \left|\innerwgn{\obsf}{\obsg}\mkern-1mu \right| \\
   &\ = \left|\innerwgn{\left(\F{\obsf}_{\mathbf{0}}\basis{\mathbf{0}}+\sum_{\bs\in \cS_{A}}\F{\obsf}_{\bs}\basis{\bs}\right)}{\left(\widehat{\obsg}_{\mathbf{0}}\basis{\mathbf{0}}+\sum_{\bt\in \cS_{A}}\widehat{\obsg}_{\bt}\basis{\bt}\right)}\right|\\
 &\ \leq \normwgn{\F{\obsf}_{\mathbf{0}}\basis{\mathbf{0}}+\sum_{\bs\in \cS_{A}}\F{\obsf}_{\bs}\basis{\bs}}\normwgn{\widehat{\obsg}_{\mathbf{0}}\basis{\mathbf{0}}+\sum_{\bt\in \cS_{A}}\widehat{\obsg}_{\bt}\basis{\bt}} \\
 &\ =\sqrt{|\F{\obsf}_{\mathbf{0}}|^{2}+\sum_{\bs\in \cS_{A}}|\F{\obsf}_{\bs}|^{2}}\sqrt{|\widehat{\obsg}_{\mathbf{0}}|^{2}+\sum_{\bt\in \cS_{A}}|\widehat{\obsg}_{\bt}|^{2}}.
\end{align*}
On the other hand, from the transitive symmetry of the observables $\obsf_{AB}$ and $\obsg_{AC}$ and using Lemma~\ref{observable-symmetry}, we get
\begin{align*}
    &\ \normwgn{\obsf}^{2}  = |\F{\obsf}_{\mathbf{0}}|^{2}+(|A|+|B|)\overline\eta(\obsf)\\
     &\  \normwgn{\obsg}^{2}  = |\widehat{\obsg}_{\mathbf{0}}|^{2}+(|A|+|C|)\overline\eta(\obsg).
\end{align*}
Thus, the following inequalities hold for $\obsf$ and $\obsg$:
\begin{align*}
    &\ \frac{|A|}{|A|+|B|}\left(\normwgn{\obsf}^{2}-|\F{\obsf}_{\mathbf{0}}|^{2}\right)\geq \sum_{\bs\in \cS_{A}}|\F{\obsf}_{\bs}|^{2},\\
   &\ \frac{|A|}{|A|+|C|}\left(\normwgn{\obsg}^{2}-|\widehat{\obsg}_{\mathbf{0}}|^{2}\right)\geq \sum_{\bs\in \cS_{A}}|\widehat{\obsg}_{\bs}|^{2}.
\end{align*}
Finally, we obtain
\begin{align*}
     \left| \innerwgn{\obsf}{\obsg} \mkern-1mu \right| 
      &\leq \sqrt{|\F{\obsf}_{\mathbf{0}}|^{2}+\sum_{\bs\in \cS_{A}}|\F{\obsf}_{\bs}|^{2}}\sqrt{|\widehat{\obsg}_{\mathbf{0}}|^{2}+\sum_{\bt\in \cS_{A}}|\widehat{\obsg}_{\bt}|^{2}}\\
     &\leq \sqrt{\kappa_1\normwgn{\obsf}^{2}
         +(1-\kappa_1)\big| \! \left\langle \obsf , \mI \right\rangle_{\rho^{\otimes n}}\mkern-4mu \big|^{2}}\\
       & \qquad \cdot    \sqrt{\kappa_2\normwgn{\obsg}^{2}
         +(1-\kappa_2)\big| \! \left\langle \obsg , \mI \right\rangle_{\rho^{\otimes n}}\mkern-4mu \big|^{2}}. \qedhere
\end{align*}
\end{proof}
\begin{rem}
Lemma~\ref{lem:corr-ineq} also holds for the KMS inner product and the same proof applies simply by changing all GNS inner products and norms to KMS inner products and norms.
\end{rem}

\end{appendices}

\printbibliography

\end{document}